\documentclass[11pt]{article}

\usepackage{epsfig,subfigure}
\usepackage{amssymb}
\usepackage[fleqn]{amsmath}
\usepackage{color}
\usepackage{arydshln}
\definecolor{orange}{rgb}{1,0.5,0}

\def\QED{\mbox{\rule[0pt]{1.5ex}{1.5ex}}}
\def\proof{\noindent\hspace{2em}{\it Proof: }}

\usepackage{graphicx}

\usepackage{amssymb}
\usepackage{amsmath,amsfonts,amssymb}
\usepackage{verbatim}
\usepackage{stfloats}
\usepackage[bookmarks=false]{}

\newtheorem{theorem}{Theorem}

\newtheorem{lemma}{Lemma}

\newcommand\blfootnote[1]{%
  \begingroup
  \renewcommand\thefootnote{}\footnote{#1}%
  \addtocounter{footnote}{-1}%
  \endgroup
}

\begin{document}
\date{}
\title{Private Information Retrieval\\ 
from MDS Coded Data with Colluding Servers: \\
Settling a Conjecture by Freij-Hollanti et al.
%\thanks{This work is supported by NSF grants CCF-1317351 and CCF-0963925.}
}
%\author{\IEEEauthorblockN{Hua Sun and Syed A. Jafar}
%\IEEEauthorblockA{Center for Pervasive Communications and Computing (CPCC)\\
%University of California Irvine, Irvine, CA 92697
%%\\ Email: \{huas2, syed\}@uci.edu
%}}
\author{ \normalsize Hua Sun and Syed A. Jafar \\
%{\small Center for Pervasive Communications and Computing (CPCC)}\\
%{\small University of California Irvine, Irvine, CA 92697}\\
%{\small \it Email: \{huas2, syed\}@uci.edu}
}

\maketitle

\blfootnote{Hua Sun (email: huas2@uci.edu) and Syed A. Jafar (email: syed@uci.edu) are with the Center of Pervasive Communications and Computing (CPCC) in the Department of Electrical Engineering and Computer Science (EECS) at the University of California Irvine. %The work is supported by grants from  ONR, NSF and ARL. The results of this work were submitted in part for presentation at IEEE ISIT 2016 and IEEE GLOBECOM 2016. 
}

\maketitle

\begin{abstract}
%To be considered for the 2016 IEEE Jack Keil Wolf ISIT Student Paper Award. 
A $(K, N, T, K_c)$ instance of the MDS-TPIR problem is comprised of $K$ messages and $N$ distributed servers. Each message is separately encoded through a $(K_c, N)$ MDS storage code. A user wishes to retrieve one message, as efficiently as possible,  while revealing no information about the desired message index to any colluding set of up to  $T$ servers. The fundamental limit on the efficiency of retrieval, i.e., the capacity of MDS-TPIR is known only at the extremes where either $T$ or $K_c$ belongs to $\{1,N\}$. The focus of this work is a recent conjecture by Freij-Hollanti, Gnilke, Hollanti and Karpuk which offers a general capacity expression for MDS-TPIR. We prove that the conjecture is false by presenting as a counterexample a PIR scheme for the setting $(K, N, T, K_c) = (2,4,2,2)$, which achieves the rate $3/5$, exceeding the conjectured capacity, $4/7$. Insights from the counterexample lead us to capacity characterizations for various instances of MDS-TPIR including all cases with $(K, N, T, K_c) = (2,N,T,N-1)$, where $N$ and $T$ can be arbitrary. 
\end{abstract}
\newpage

\section{Introduction}
Private Information Retrieval (PIR) is the problem of retrieving one out of $K$ messages from $N$ distributed servers (each stores all $K$ messages) in such a way that any individual server learns no information about which message is being retrieved. The rate of a PIR scheme is the ratio of the number of bits of the desired message to the total number of bits downloaded from all servers. The supremum of achievable rates is the capacity of PIR. The capacity of PIR was shown in \cite{Sun_Jafar_PIR} to be 
\begin{eqnarray}
C_{\mbox{\scriptsize PIR}}&=&\left(1+\frac{1}{N}+\frac{1}{N^2}+\cdots+\frac{1}{N^{K-1}}\right)^{-1} \label{eq:CPIR}
\end{eqnarray}
The capacity of several variants of PIR has also since been characterized in \cite{Sun_Jafar_PIR, Sun_Jafar_TPIR, Sun_Jafar_SPIR, Sun_Jafar_MPIR, Banawan_Ulukus}.

The focus of this work is on a recent conjecture by Freij-Hollanti, Gnilke, Hollanti and Karpuk (FGHK conjecture, in short) in \cite{FREIJ_HOLLANTI} which offers a  capacity expression for a generalized form of PIR, called MDS-TPIR. MDS-TPIR involves two additional parameters: $K_c$ and $T$, which generalize the storage and privacy constraints, respectively. Instead of replication, each message is  encoded through a $(K_c, N)$ MDS storage code, so that the information stored at any $K_c$ servers is exactly enough to recover all $K$ messages. Privacy must be preserved not just from each individual server, but from any colluding set of up to  $T$ servers. MDS-TPIR is a generalization of PIR, because setting both $T=1$ and $K_c=1$ reduces MDS-TPIR to the original PIR problem for which the capacity is already known (see (\ref{eq:CPIR})).

The capacity of MDS-TPIR is known only at the degenerate extremes -- when either $T$ or $K_c$ takes the value $1$ or $N$. If either $T$ or $K_c$ is equal to $N$ then by analogy to the single server setting it follows immediately that the user must download all messages, i.e., the capacity is $1/K$. If $K_c=1$ or $T=1$, then the problem specializes to TPIR, and  MDS-PIR, respectively.  The capacity of TPIR ($K_c=1$) was shown in \cite{Sun_Jafar_TPIR} to be
\begin{eqnarray}
C_{\mbox{\scriptsize TPIR}}&=&\left(1+\frac{T}{N}+\frac{T^2}{N^2}+\cdots+\frac{T^{K-1}}{N^{K-1}}\right)^{-1} 
\end{eqnarray}
The capacity of MDS-PIR ($T=1$) was characterized by Banawan and Ulukus in \cite{Banawan_Ulukus},  as
\begin{eqnarray}
C_{\mbox{\scriptsize MDS-PIR}}&=&\left(1+\frac{K_c}{N}+\frac{K_c^2}{N^2}+\cdots+\frac{K_c^{K-1}}{N^{K-1}}\right)^{-1} 
\end{eqnarray}
It is notable that $K_c$ and $T$  play similar roles in the two capacity expressions.

The capacity achieving scheme of Banawan and Ulukus \cite{Banawan_Ulukus} improved upon a scheme proposed earlier by Tajeddine and Rouayheb in \cite{Tajeddine_Rouayheb}. 
Tajeddine and Rouayheb also proposed an achievable scheme for  MDS-TPIR for  the $T=2$ setting. The scheme was generalized by Freij-Hollanti et al. \cite{FREIJ_HOLLANTI} to the  $(K, N, T, K_c)$ setting, $T+K_c\leq N$, where it achieves the rate $1 - \frac{T+K_c -1}{N}$. Remarkably, the rate achieved by this scheme does not depend on the number of messages, $K$. In support of the plausible asymptotic ($K\rightarrow\infty$) optimality of their scheme, and based on the intuition from existing capacity expressions for PIR, MDS-PIR and TPIR, Freij-Hollanti et al. conjecture that if $T+K_c\leq N$, then the capacity of MDS-TPIR is given by the following expression.

\vspace{0.05in}

\noindent{\sc FGHK Conjecture \cite{FREIJ_HOLLANTI}}:
\begin{eqnarray}
C^{\mbox{\scriptsize conj}}_{\mbox{\scriptsize MDS-TPIR}} =\left( 1 + \frac{T+K_c -1}{N} + \cdots + \frac{(T+K_c -1)^{K-1}}{N^{K-1}} \right)^{-1}
\end{eqnarray}
The conjecture is appealing for its generality and elegance as it captures all four parameters, $K, N, T, K_c$ in a compact form. $T$ and $K_c$ appear as interchangeable terms,  and the capacity expression appears to be a natural extension of the  capacity expressions for TPIR and MDS-PIR. Indeed, the conjectured capacity recovers the known capacity of TPIR if we set $K_c=1$ and that of MDS-PIR if we set $T=1$. However, in all non-degenerate cases where $T,K_c\notin\{1,N\}$, the capacity of MDS-TPIR, and therefore the validity of the conjecture is unknown. In fact, in all these cases the problem is open on \emph{both} sides, i.e., the conjectured capacity expression is neither known to be achievable, nor known to be an outer bound. The lack of any non-trivial outer bounds for MDS-TPIR is also recently highlighted in \cite{Kumar_Rosnes_Amat}. This intriguing combination of plausibility, uncertainty and generality of the FGHK conjecture motivates our work. Our  contribution is summarized next.

\subsubsection*{Summary of Contribution}
As the main outcome of this work, we disprove the FGHK conjecture. For our counterexample, we consider the setting $(K, N, T, K_c) = (2, 4, 2, 2)$ where the data is stored using the $(2,4)$  MDS code $(x,y)\rightarrow (x,y,x+y,x+2y)$. The conjectured capacity for this setting is $4/7$. We show that the rate $3/5 > 4/7$ is achievable, thus disproving the conjecture. As a converse argument, we show that no  (scalar or vector) linear PIR scheme can achieve a rate higher than $3/5$ for this MDS storage code subject to $T=2$ privacy. %Furthermore, no PIR scheme (linear or non-linear)  can achieve a rate higher than $8/13$ for the same  $(K, N, T, K_c) = (2, 4, 2, 2)$ setting. Remarkably, $8/13$ is shown to be the capacity if the colluding sets of servers are restricted to the class of cyclically adjacent pairs, i.e., servers $\{1,2\}, \{2,3\}, \{3,4\}, \{4,1\}$. However, unlike TPIR, where if $N=mT$, $m\in\mathbb{Z}_+$, the capacity remains unchanged even if colluding sets are restricted to servers $\{1,\cdots, T\}, \{T+1, \cdots, 2T\}, \cdots,\{(m-1)T+1, \cdots, N\}$, we show that such a restriction can strictly increase capacity for MDS-TPIR.

The insights from the counterexample lead us to characterize the exact capacity of various instances of MDS-TPIR. This includes all cases with $(K,N,T,K_c)=(2,N,T,N-1)$, where $N$ and $T$ can be arbitrary. The capacity for these cases turns out to be 
\begin{eqnarray}
C&=&\frac{N^2-N}{2N^2-3N+T}
\end{eqnarray}
Note that this is the information theoretic capacity, i.e., for $K=2$ messages, no $(N-1,N)$ MDS storage code and no PIR scheme (linear or non-linear) can beat this rate, which is achievable with the simple MDS storage code $(x_1, x_2,\cdots, x_{N-1})\rightarrow (x_1, x_2, \cdots, x_{N-1}, \sum_{i=1}^{N-1} x_i)$ and a linear PIR scheme. 

The general capacity expression for MDS-TPIR remains unknown. However, we are able to show that  it cannot be symmetric in $K_c$ and $T$, i.e., the two parameters are not interchangeable in general. Also, between $K_c$ and $T$ the capacity expression does not consistently favor one over the other. These findings are illustrated by the following four cases for which the capacity is settled.
\begin{eqnarray*}
\begin{array}{c|c|c|c|c|}
&\multicolumn{4}{c}{(K,N,T,K_c)}\vline\\ \cline{2-5}
&(2,4,2,3)&(2,4,3,2)&(2,4,1,3)&(2,4,3,1)\\\hline
\mbox{Capacity}&6/11&4/7&4/7&4/7\\\hline
\mbox{Ref.}&\mbox{Theorem \ref{thm:class}}&\mbox{Section \ref{sec:Ex2}}&\cite{Banawan_Ulukus}&\cite{Sun_Jafar_TPIR}\\\hline
\end{array}
\end{eqnarray*}
The first two columns show that the capacity is not symmetric in $K_c$ and $T$, since switching their values changes the capacity. The first two columns also suggest that increasing $K_c$ hurts capacity more than increasing $T$. However, considering columns $3$ and $4$ as the baseline where the capacities are equal, and comparing the drop in capacity from column $3$ to column $1$ when $T$ is increased, versus no change in capacity from column $4$ to column $2$ when $K_c$ is increased shows the opposite trend. Therefore, neither $T$ nor $K_c$ is consistently dominant in terms of the sensitivity of capacity to these two parameters. 

Finally,  taking an asymptotic view of capacity of MDS-TPIR, we show that  if $T+K_c>N$, then the capacity collapses to $0$ as the number of messages $K\rightarrow\infty$. This is consistent with the restriction of $T+K_c\leq N$  that is required by the achievable scheme of Freij-Hollanti et al. whose  rate does not depend on $K$. %In general, however, even the asymptotic optimality of their scheme, as $K\rightarrow\infty$, remains unknown.

{\it Notation:} For $n_1, n_2 \in \mathbb{Z}$, define the notation $[n_1: n_2]$ as the set $\{n_1, n_1+1,\cdots, n_2\}$, $A_{n_1:n_2}$ as the vector $(A_{n_1}, A_{n_1+1}, \cdots, A_{n_2})$, and  $S(n_1 : n_2,:)$ as the submatrix of a matrix $S$ formed by retaining only the $n_1^{th}$ to the $n_2^{th}$ rows. The notation $X \sim Y$ is used to indicate that $X$ and $Y$ are identically distributed. The cardinality of a set $\mathcal{I}$ is denoted as $|\mathcal{I}|$. The determinant of a matrix $S$ is denoted as $|S|$. For an index set $\mathcal{I} = \{i_1, \cdots, i_{n}\}$ such that $i_1 < \cdots < i_{n}$, the notation $A_{\mathcal{I}}$ represents the vector $(A_{i_1}, \cdots, A_{i_{n}})$. $(V_1; V_2; \cdots; V_n)$ refers to a matrix whose $i^{th}$ row vector is $V_i, i \in [1:n]$.

\section{Problem Statement}\label{sec:model}
Consider\footnote{While the problem statement is presented in its general form, we will primarily consider cases with $K=2$ messages in this paper (outer bounds for larger $K$ are presented in Section \ref{sec:converse}).} $K$ independent messages $W_1, \cdots, W_K\in\mathbb{F}_p^{L\times 1}$, each represented as an $L\times 1$ vector comprised of  $L$ i.i.d. uniform symbols from a finite field $\mathbb{F}_p$ for a prime $p$. In $p$-ary units,
\begin{eqnarray}
H(W_1) &=& \cdots = H(W_K) = L \label{h2}\\
H(W_1, \cdots, W_K) &=& H(W_1) + \cdots + H(W_K) 
\end{eqnarray}
There are $N$ servers. The $n^{th}$ server stores $(W_{1n}, W_{2n}, \cdots, W_{Kn})$, where $W_{kn}\in\mathbb{F}^{\frac{L}{K_c}\times 1}$ represents $L/K_c$ symbols from $W_k, k \in [1:K]$. 
\begin{eqnarray}
H(W_{kn} | W_k) = 0, ~H(W_{kn}) = L/K_c \label{storage_size}
\end{eqnarray}
We require the storage system to satisfy the MDS property, i.e., from the information stored in any $K_c$ servers, we can recover each message, i.e.,
\allowdisplaybreaks
\begin{eqnarray}
[\mbox{MDS}] ~ H(W_k | W_{k \mathcal{K}_c}) = 0, \forall \mathcal{K}_c \subset [1:N], |\mathcal{K}_c| = K_c \label{mds_property}
\end{eqnarray}
Let us use $\mathcal{F}$ to denote a random variable privately generated by the user, whose realization is not available to the servers. $\mathcal{F}$ represents the randomness in the strategies followed by the user. Similarly, $\mathcal{G}$ is a random variable that determines the random strategies followed by the servers, and whose realizations are assumed to be known to all the servers and to the user. The user privately generates $\theta$ uniformly from $[1:K]$ and wishes to retrieve $W_\theta$ while keeping $\theta$ a secret from each server. $\mathcal{F}$ and $\mathcal{G}$ are generated independently and before the realizations of the messages or the desired message index are known, so that
\begin{eqnarray}
H(\theta, \mathcal{F}, \mathcal{G}, W_1, \cdots, W_K)  = H(\theta) + H(\mathcal{F}) + H(\mathcal{G}) + H(W_1) + \cdots + H(W_K) \label{indep}
\end{eqnarray}
Suppose $\theta = k$. In order to retrieve $W_k, k \in [1:K]$ privately, the user privately generates $N$ random queries, $Q_1^{[k]}, \cdots, Q_N^{[k]}$.
\begin{eqnarray}
H(Q_1^{[k]}, \cdots, Q_N^{[k]} | \mathcal{F}) = 0, \forall k \in [1:K] \label{query_det}
\end{eqnarray}
The user sends query $Q_n^{[k]}$ to the $n^{th}$ server, $n \in [1:N]$. Upon receiving $Q_n^{[k]}$, the $n^{th}$ server generates an answering string $A_n^{[k]}$, which is a function of the received query $Q_n^{[k]}$, the stored information $W_{1n}, \cdots, W_{Kn}$ and $\mathcal{G}$,
\begin{eqnarray}
H(A_n^{[k]} | Q_n^{[k]}, W_{1n}, \cdots, W_{Kn}, \mathcal{G}) = 0 \label{answer_det}
\end{eqnarray}
Each server returns to the user its answer $A_n^{[k]}$.\footnote{If the $A_n^{[k]}$ are obtained as inner products of query vectors and stored message vectors, then such a PIR scheme is called a linear PIR scheme.}

From all the information that is now available at the user $(A_{1:N}^{[k]}, Q_{1:N}^{[k]}, \mathcal{F},\mathcal{G})$, the user decodes the desired message $W_k$ according to a decoding rule that is specified by the PIR scheme. Let $P_e$ denote the probability of error achieved with the specified decoding rule.

To protect the user's privacy, the $K$ strategies must be indistinguishable (identically distributed) from the perspective of any subset $\mathcal{T} \subset [1:N]$ of at most $T$ colluding servers, i.e., the following privacy constraint must be satisfied.
\begin{eqnarray}
[\mbox{$T$-Privacy}]~ 
(Q_{\mathcal{T}}^{[k]}, A_{\mathcal{T}}^{[k]}, \mathcal{G}, W_{1\mathcal{T}}, \cdots, W_{K\mathcal{T}}) \sim (Q_{\mathcal{T}}^{[k']}, A_{\mathcal{T}}^{[k']}, \mathcal{G}, W_{1\mathcal{T}}, \cdots, W_{K\mathcal{T}}), \notag \\
 \forall k, k' \in [1:K], \forall \mathcal{T} \subset [1:N], |\mathcal{T}| = T \label{privacy}
\end{eqnarray}

The PIR rate characterizes how many bits of desired information are retrieved per downloaded bit and is defined as follows.
\begin{eqnarray}
R = L/D
\end{eqnarray}
where $D$ is the expected value of the total number of bits downloaded by the user from all the servers. 

A rate $R$ is said to be $\epsilon$-error achievable if there exists a sequence of PIR schemes, indexed by $L$, each of rate greater than or equal to $R$, for which $P_e \rightarrow 0$ as $L \rightarrow \infty$. Note that for such a sequence of PIR schemes, from Fano's inequality, we must have
\begin{eqnarray}
[\mbox{Correctness}]~
 o(L) &=& \frac{1}{L} H(W_k|A_{1:N}^{[k]}, Q_{1:N}^{[k]}, \mathcal{F}, \mathcal{G})  \\
&\overset{(\ref{query_det})}{=}& \frac{1}{L} H(W_k|A_{1:N}^{[k]}, \mathcal{F}, \mathcal{G}) \label{corr}
\end{eqnarray}
where $o(L)$ represents a term whose value approaches zero as $L$ approaches infinity. The supremum of $\epsilon$-error achievable rates is called the capacity $C$.\footnote{Alternatively, the capacity may be defined with respect to zero error criterion, i.e., the supreme of zero error achievable rates where a rate $R$ is said to be zero error achievable if there exists (for some $L$) a PIR scheme of rate greater than or equal to $R$ for which $P_e = 0$. }

\section{Settling the Conjecture}\label{sec:main}
Our main result, which settles the FGHK conjecture, is stated in the following theorem.
\begin{theorem}\label{thm:disprove}
For the MDS-TPIR problem with $K = 2$ messages, $N = 4$ servers, $T = 2$ privacy and the $(K_c, N) = (2,4)$ MDS storage code  $(x,y)\rightarrow (x,y,x+y,x+2y)$, a rate of $3/5$ is achievable. Since the achievable rate exceeds the conjectured capacity of $4/7$ for this setting, the FGHK conjecture is false.
\end{theorem}

{\it Proof:} We present a scheme that achieves rate $3/5$. We assume that each message is comprised of $L=12$  symbols from $\mathbb{F}_p$ for a  sufficiently\footnote{{\color{black} It suffices to choose $p=349$ for Theorem \ref{thm:disprove}. In general, the appeal to large field size, analogous to the random coding argument in information theory, is made to prove the existence of a scheme, but may not be essential to the construction of the PIR scheme. To underscore this point, Section \ref{sec:ex_small_field} includes some examples of MDS-TPIR capacity achieving schemes over small fields.  }} large prime $p$. Define ${\bf a}\in\mathbb{F}_p^{6\times1}$ as the $6\times 1$ vector $(a_1;a_2;\cdots; a_6)$ comprised of i.i.d. uniform symbols $a_i\in\mathbb{F}_p$. Vectors ${\bf b}, {\bf c}, {\bf d}$ are defined similarly. Messages $W_1, W_2$ are defined in terms of these vectors as follows.
\begin{eqnarray}
W_1 = ({\bf a}; {\bf b}) && W_2 = ({\bf c}; {\bf d})
\end{eqnarray}
\subsection{Storage Code}
The storage is specified as
\begin{eqnarray}
(W_{11},W_{12}, W_{13}, W_{14}) &=& ({\bf a}, {\bf b}, {\bf a +b}, {\bf a+2b})\label{eq:mds1}\\
(W_{21},W_{22}, W_{23}, W_{24}) &=& ({\bf c}, {\bf d}, {\bf c +d}, {\bf c+2d})\label{eq:mds2}
\end{eqnarray}
Recall that $W_{kn}$ is the information about message $W_k$ that is stored at Server $n$. Thus, Server $1$ stores $({\bf a,c})$, Server $2$ stores $({\bf b,d})$, Server $3$ stores $({\bf a+b, c+d})$, and Server $4$ stores $({\bf a+2b, c+2d})$. In particular, each server stores $6$ symbols for each message, for a total of $12$ symbols per server. Any two servers store just enough information to recover both messages, thus the MDS storage criterion is satisfied.

\subsection{Construction of Queries}
The query to each server $Q_n^{[k]}$ is comprised of two parts,  denoted as $Q_n^{[k]}(W_1), Q_n^{[k]}(W_2)$. Each part contains $3$ row vectors, also called query vectors, along which the server should project its corresponding stored message symbols.
\begin{eqnarray}
Q_n^{[k]} = (Q_n^{[k]}(W_1), Q_n^{[k]}(W_2)) \label{query_2}
\end{eqnarray}
In preparation for the construction of the queries, let us denote the set of all full rank $6 \times 6$ matrices over $\mathbb{F}_p$ as $\mathcal{S}$. The user privately chooses two matrices, $S$ and $S'$, independently and uniformly from $\mathcal{S}$. Label the rows of $S$ as $V_1, V_2, V_3, V_4, V_5, V_6$, and the rows of $S'$ as $U_0, U_1, U_2, U_3, U_4, U_5$. Define
\begin{eqnarray}
\mathcal{V}_1=\{V_1, V_2, V_3\},&&\mathcal{U}_1=\{U_0, U_6, U_8\}\\
\mathcal{V}_2=\{V_1, V_4, V_5\},&&\mathcal{U}_2=\{U_0, U_7, U_9\}\\
\mathcal{V}_3=\{V_2, V_4, V_6\},&&\mathcal{U}_3=\{U_0, U_1, U_3\}\\
\mathcal{V}_4=\{V_3, V_5, V_6\},&&\mathcal{U}_4=\{U_0, U_2, U_4\}
\end{eqnarray}
 $U_6, U_7, U_8, U_9$ are obtained as follows.
\begin{eqnarray}
U_6 = U_1 + U_2, &&U_7 = U_1 + 2U_2 \label{u1}\\
U_8 = U_3 + U_4, &&U_9 = U_3 + 2U_4 \label{u2}
\end{eqnarray}

As a preview of what we are trying to accomplish, we note that for Server $n\in[1:4]$,   $\mathcal{V}_n$ will be used as the query vectors for desired message symbols, while $\mathcal{U}_n$ will be used as query vectors for undesired message symbols. Since $K_c=2$, the same query vector $V_i$ sent to two different servers will recover $2$ independent desired symbols.  Each $V_i, i\in[1:6]$, is used exactly twice, so all queries for desired symbols will return independent information for a total of $12$ independent desired symbols. On the other hand, for undesired symbols note that $U_0$ is used as the query vector to all $4$ servers, but because $K_c=2$, it can only produce $2$ independent symbols, i.e., $2$ of the $4$ symbols are redundant. The dependencies introduced via (\ref{u1}),(\ref{u2}) are carefully chosen to  ensure that the queries along $U_1, U_2, U_6, U_7$ will produce only  $3$ independent symbols. Similarly, the queries along $U_3,U_4, U_8,U_9$ will produce only $3$ independent symbols. Thus, all the queries for the undesired message will produce a total of only $8$ independent symbols. The $12$ independent desired symbols and $8$ independent undesired symbols will be resolved from a total of $12+8=20$ downloaded symbols, to achieve the rate $12/20=3/5$. To ensure $T=2$ privacy, the $\mathcal{U}_i$ and $\mathcal{V}_i$ queries will be made indistinguishable from the perspective of any $2$ colluding servers. The key to the $T=2$ privacy is that any $\mathcal{V}_n, \mathcal{V}_{n'}$, $n\neq n'$ have one element in common. Similarly,  any $\mathcal{U}_n, \mathcal{U}_{n'}$, $n\neq n'$ also have one element in common. This is a critical aspect of the construction.

Next we provide a detailed description of the queries and downloads  for message $W_k, k \in [1:2]$, both when $W_k$ is desired and when it is not desired. To simplify the notation, we will denote $W_k = ({\bf x}; {\bf y})$. Note that when $k = 1$, $({\bf x}; {\bf y}) = ({\bf a}; {\bf b})$ and when $k = 2$, $({\bf x}; {\bf y}) = ({\bf c}; {\bf d})$.

\subsubsection{\it Case 1. $W_k$ is Desired}
The query sent to Server $n$ is a $3\times 6$ matrix whose rows are the $3$ vectors in $\mathcal{V}_n$. The ordering of the rows is uniformly random, i.e.,
\begin{eqnarray}
{\mbox{Server $n$}}:~Q_n^{[k]}(W_k) = \mathbb{\pi}_n (\mathcal{V}_n),&& n\in[1:4]
\end{eqnarray}
For a set $\mathcal{V}=\{V_{i_1},V_{i_2},V_{i_3}\}$, $\mathbb{\pi}_n (\mathcal{V})$ is equally likely to return any one of the $6$ possibilities: $(V_{i_1}; V_{i_2}; V_{i_3})$, $(V_{i_1}; V_{i_3}; V_{i_2})$, $(V_{i_2}; V_{i_1}; V_{i_3})$, $(V_{i_2}; V_{i_3}; V_{i_1})$, $(V_{i_3}; V_{i_1}; V_{i_2})$ and $(V_{i_3}; V_{i_2}; V_{i_1})$. The $\pi_n$ are independently chosen for each $n\in[1:4]$.

After receiving the $3$ query vectors $Q_n^{[k]}(W_k)$,  Server $n$ projects its stored $W_{kn}$ symbols along these vectors.  
This creates three linear combinations of $W_{kn}$ symbols (denoted as $A_n^{[k]}(W_k)$).
\begin{eqnarray}
A_n^{[k]}(W_k) = Q_n^{[k]}(W_k) W_{kn}
\end{eqnarray}
Define $k^c=3-k$ as the complement of $k$, i.e., $k^c=1$ if $k=2$ and vice versa. The answers $A_n^{[k]}$ to be sent to the user will be constructed eventually by combining $A_n^{[k]}(W_k)$ and $A_n^{[k]}(W_{k^c})$, since separately sending these answers will be too inefficient. The details of this combining process will be specified later. Next we note an important property of the construction.

{\it Desired Symbols Are Independent:} We show that if the user can recover $A_{1:4}^{[{k}]}(W_k)$ from the downloads, then he can recover all $12$ symbols of $W_k$. From $A_{1:4}^{[k]}(W_k)$ the user recovers the $12$ symbols $V_1{\bf x}$, $V_2{\bf x}$, $V_3{\bf x}$, $V_1{\bf y}$, $V_4{\bf y}$, $V_5{\bf y}$, $V_2({\bf x+y})$, $V_4({\bf x+y})$, $V_6({\bf x+y})$, $V_3({\bf x+2y})$, $V_5({\bf x+2y})$, $V_6({\bf x+2y})$. From these $12$ symbols, he recovers $V_i{\bf x}$ and $V_i{\bf y}$ for all $i\in[1:6]$.
Since $S = (V_1; V_2; V_3; V_4; V_5; V_6)$ has full rank (invertible) {\color{black} and the user knows $V_{1:6}$}, he recovers all symbols in ${\bf x}$ and ${\bf y}$ (thus $W_k$).

\subsubsection{\it Case 2. $W_k$ is Undesired}
Similarly, the query sent to Server $n$ is a $3\times 6$ matrix whose rows are the $3$ vectors in $\mathcal{U}_n$.  The ordering of the rows is uniformly random for each $n$, and independent across all $n\in[1:4]$.
\begin{eqnarray}
{\mbox{Server $n$}}:~Q_n^{[k^c]}(W_k) = \mathbb{\pi}_n'(\mathcal{U}_n),&& n\in[1:4]
\end{eqnarray}
Each server projects its stored $W_{kn}$ symbols along the $3$ query vectors to obtain,
\begin{eqnarray}
A_n^{[k^c]}(W_k) = Q_n^{[k^c]}(W_k) W_{kn}
\end{eqnarray}
{\it Interfering Symbols Have Dimension $8$:}
 $A_{1:4}^{[k^c]}(W_k)$ is comprised of  $U_0{\bf x}$, $U_6{\bf x}$, $U_8{\bf x}$, $U_0{\bf y}$, $U_7{\bf y}$, $U_9{\bf y}$, $U_0({\bf x+y})$, $U_1({\bf x+y})$, $U_3({\bf x+y})$, $U_0({\bf x+2y})$, $U_2({\bf x+2y})$, $U_4({\bf x+2y})$. We now show that these $12$ symbols are dependent and have dimension only $8$.\footnote{Equivalently, the joint entropy of these $12$ variables, conditioned on $U_{0:9}$ is only $8$ $p$-ary units.} Because of (\ref{u1}) and (\ref{u2}), we have
\begin{eqnarray}
U_0 {\bf x} + U_0 {\bf y} &=& U_0 ({\bf x}+{\bf y}) \notag\\
U_0 {\bf x} + 2U_0 {\bf y} &=& U_0  ({\bf x}+2{\bf y}) \notag\\
U_6 {\bf x} + U_7 {\bf y} - U_1 ({\bf x}+{\bf y}) &=& U_2  ({\bf x}+2{\bf y}) \notag\\
U_8 {\bf x} + U_9 {\bf y} - U_3  ({\bf x}+{\bf y}) &=& U_4  ({\bf x}+2{\bf y}) \label{eq:dependent}
\end{eqnarray}
Thus, of the $12$ symbols recovered from $A_{1:4}^{[k^c]}(W_k)$, at least $4$ are linear combinations of the remaining $8$. It follows that $A_{1:4}^{[k^c]}(W_k)$ contains no more than $8$ dimensions. The number of dimensions is also not less than $8$ because, the following $8$ undesired symbols (two symbols from each server) are  independent, 
\begin{eqnarray}
{\mbox{Server 1}}:~&& U_0 {\bf x}, U_6 {\bf x} = (U_1+U_2) {\bf x} \notag\\
{\mbox{Server 2}}:~&& U_0 {\bf y}, U_9 {\bf y} = (U_3 + 2U_4) {\bf y} \notag\\
{\mbox{Server 3}}:~&& U_1 ({\bf x} + {\bf y}), U_3 ({\bf x} + {\bf y}) \notag\\
{\mbox{Server 4}}:~&& U_2 ({\bf x} + 2{\bf y}), U_4 ({\bf x} + 2{\bf y}) \label{eq:8independent}
\end{eqnarray}
To see that the 8 symbols are  independent, we add 4 new symbols ($U_1 {\bf x}$, $U_3 {\bf y}$, $U_5{\bf x}$, $U_5{\bf y}$) such that from the 12 symbols, we can recover all 12 undesired symbols ($S' {\bf x}$, $S' {\bf y}$). Since the $4$ new symbols cannot contribute more than 4 dimensions, the original 8 symbols must occupy at least 8 dimensions.

\subsection{Combining Answers for Efficient Download}\label{sec:comb_spec}
Based on the queries, each server has $3$ linear combinations of symbols of $W_1$ in  $A_n^{[k]}(W_1)$ and $3$ linear combinations of symbols of $W_2$ in $A_n^{[k]}(W_2)$ for a total of $12$ linear combinations of desired symbols and $12$ linear combinations of undesired symbols across all servers. However, recall that there are only $8$ independent linear combinations of undesired symbols. This is a fact that can be exploited to improve the efficiency of download. Specifically, we will combine the $6$ queried symbols (i.e., the $6$ linear combinations) from each server into $5$  symbols to be downloaded by the user. Intuitively, $5$ symbols from each server will give the user a total of $20$ symbols, from which he can resolve the $12$ desired and $8$ undesired symbols. 
 
The following function maps $6$ queried symbols to $5$ downloaded symbols.
\begin{eqnarray}
{\color{black}\mathcal{L}}(X_{1}, X_2, X_3, Y_1, Y_2, Y_3) &= (X_1, X_2, Y_1, Y_2, X_3 + Y_3)
\end{eqnarray}
Note that the first four symbols are directly downloaded and only the last symbol is mixed. The desired and undesired symbols are combined to produce the answers as follows.
\begin{eqnarray}
A_n^{[k]} = \mathcal{L}(C_n A_n^{[k]}(W_1), C_n A_n^{[k]}(W_{2})) 
\end{eqnarray}
where $C_n$ are deterministic $3\times 3$ matrices, that are required to satisfy the following two properties. Denote the first $2$ rows of $C_n$ as $\overline{C}_n$.
\begin{enumerate}
\item[{\it P1.}] All $C_n$ must have full rank.
\item[{\it P2.}] {\color{black} For all $(3!)^4$ distinct realizations of $\pi_n', n \in [1:4]$,} the $8$  linear combinations of the undesired message symbols that are directly downloaded ($2$ from each server), $\overline{C}_{1} A_1^{[k]}(W_{k^c})$, $\overline{C}_{2} A_2^{[k]}(W_{k^c})$, $\overline{C}_{3} A_3^{[k]}(W_{k^c})$, $\overline{C}_{4} A_4^{[k]}(W_{k^c})$ are  independent. 
\end{enumerate}
As we will prove in the sequel, it is not difficult to find matrices that satisfy these properties. In fact, these properties are `generic', i.e., uniformly random choices of $C_n$  matrices will satisfy these properties with probability approaching $1$ as the field size approaches infinity. The appeal to generic property will be particularly useful as we consider larger classes of MDS-TPIR settings. Those (weaker) proofs apply here as well. However, for the particular setting of Theorem \ref{thm:disprove}, based on a brute force search we are able to strengthen the proof by presenting the following explicit choice of $C_n, n\in[1:4]$ which satisfies both properties over $\mathbb{F}_{349}$.

\begin{eqnarray}
C_1 = \left(
\begin{array}{ccc}
1 & 2 & 3\\
6 & 5 & 4\\
0 & 0 & 1
\end{array} \right),~
C_2 = \left(
\begin{array}{ccc}
1 & 7 & 3\\
11 & 9 & 8\\
0 & 0 & 1
\end{array} \right),~
C_3 = \left(
\begin{array}{ccc}
1 & 10 & 8\\
7 & 5 & 4\\
0 & 0 & 1
\end{array} \right),~
C_4 = \left(
\begin{array}{ccc}
1 & 3 & 5\\
12 & 9 & 3\\
0 & 0 & 1
\end{array} \right) 
\end{eqnarray}
Property $P1$ is trivially verified. Property $P2$ is  verified by considering one by one, all of the $6^4$ distinct realizations of $\pi_n', n\in[1:4]$. To show how this is done, let us consider one case here. Suppose the realization of the permutations is such that
\begin{eqnarray}
\pi_1'(\mathcal{U}_1) &=& (U_0, U_6, U_8)\\
\pi_2'(\mathcal{U}_2) &=& (U_0, U_9, U_7)\\ 
\pi_3'(\mathcal{U}_3) &=& (U_1, U_3, U_0)\\
\pi_4'(\mathcal{U}_4) &=& (U_2, U_4, U_0)
\end{eqnarray}
then we have
\begin{eqnarray}
(\overline{C}_{1} A_1^{[k]}(W_{k^c}); \cdots; \overline{C}_{4} A_4^{[k]}(W_{k^c})) = 
\underbrace{\left(\begin{array}{cccccccc}
1 & 2 & 0 & -3 & 0 & 3 & 0 & 3\\
6 & 5 & 0 & -4 & 0 & 4 & 0 & 4\\
0 & -3 & 1 & 7 & 3 & 0 & 3 & 0\\
0 & -8 & 11 & 9 & 8 & 0 & 8 & 0\\
8 & 0 & 8 & 0 & 1 & 10 & 0 & 0\\
4 & 0 & 4 & 0 & 7 & 5 & 0 & 0\\
5 & 0 & 10 & 0 & 0 & 0 & 1 & 3\\
3 & 0 & 6 & 0 & 0 & 0 & 12 & 9
\end{array}
\right)}_{\triangleq \mathcal{C}}
\left( \begin{array}{c}
U_0 {\bf x} \\
U_6 {\bf x} \\ 
U_0 {\bf y} \\ 
U_9 {\bf y} \\ 
U_1  ({\bf x}+{\bf y})\\ 
U_3  ({\bf x}+{\bf y})\\
U_2  ({\bf x}+2{\bf y})\\
U_4  ({\bf x}+2{\bf y})
\end{array}
\right)\label{eq:noU}
\end{eqnarray}
The determinant of $\mathcal{C}$  over $\mathbb{F}_{349}$ is $321$. Since the determinant is non-zero, all of its $8$ rows are linearly independent.  Note that the test for property $P2$ does not depend on the realizations of $U_i$ vectors. To see why this is true, note that the $8$ linear combinations of $({\bf x, y})$ in the rightmost column vector of (\ref{eq:noU}) are linearly independent. Therefore, if $\mathcal{C}$ is an invertible matrix then the $8$ directly downloaded linear combinations on the LHS of (\ref{eq:noU}) are also  independent (have joint entropy $8$ $p$-ary units, conditioned on $U_{0:9}$).

At this point the construction of the scheme is complete. All that remains now is to prove that the  scheme is correct, i.e., it retrieves the desired message,  and that it is $T=2$ private.

\subsection{\it The Scheme is Correct (Retrieves Desired Message)}
As noted previously, the first $4$ variables in the output of the $\mathcal{L}$ function are obtained directly, i.e., $\overline{C}_{1} A_1^{[k]}(W_1)$, $\overline{C}_{2} A_2^{[k]}(W_1)$, $\overline{C}_{3} A_3^{[k]}(W_1)$, $\overline{C}_{4} A_4^{[k]}(W_1)$ and $\overline{C}_{1} A_1^{[k]}(W_2)$, $\overline{C}_{2} A_2^{[k]}(W_2)$, $\overline{C}_{3} A_3^{[k]}(W_2)$, $\overline{C}_{4} A_4^{[k]}(W_2)$ are all directly recovered. By property {\it P2} of $C_n$, $\overline{C}_{1} A_1^{[k]}(W_{k^c})$, $\overline{C}_{2} A_2^{[k]}(W_{k^c})$, $\overline{C}_{3} A_3^{[k]}(W_{k^c})$, $\overline{C}_{4} A_4^{[k]}(W_{k^c})$ are linearly independent. Since the user  has recovered $8$ independent dimensions of interference, and interference only spans $8$ dimensions, all interference is recovered and eliminated. Once the interference is eliminated, since $C_n$ matrices have full rank, the user is left with $12$ independent linear combinations of desired symbols, from which he is able to recover the $12$ desired message symbols. Therefore the scheme is correct.

\subsection{\it The Scheme is Private (to any $T=2$ Colluding Servers)}
To prove that the scheme is $T = 2$ private (refer to (\ref{privacy})), it suffices to show that the queries for any $2$ servers are identically distributed, regardless of which message is desired. Since each query is made up of two independently generated parts, one for each message, it suffices to prove that the query vectors for a message (say $W_k$) are identically distributed, regardless of whether the message is desired or undesired,
\begin{eqnarray}
\left(Q_{n_1}^{[k]}(W_k), Q_{n_2}^{[k]}(W_k)\right) \sim \left(Q_{n_1}^{[k^c]}(W_k), Q_{n_2}^{[k^c]} (W_k)\right), 
~\forall n_1, n_2  \in [1:4], n_1 < n_2 \label{space_privacy}
\end{eqnarray}
Note that  
\begin{eqnarray}
\left(Q_{n_1}^{[k]}(W_k), Q_{n_2}^{[k]}(W_k)\right) = \left(\mathbb{\pi}_{n_1}(\mathcal{V}_{n_1}), \mathbb{\pi}_{n_2}(\mathcal{V}_{n_2}) \right) \\
\left(Q_{n_1}^{[k^c]}(W_k), Q_{n_2}^{[k^c]}(W_k)\right) = \left(\mathbb{\pi}_{n_1}' (\mathcal{U}_{n_1}), \mathbb{\pi}_{n_2}' (\mathcal{U}_{n_2}) \right)
\end{eqnarray}
Therefore, to prove (\ref{space_privacy}) it suffices to show the following.
\begin{eqnarray}
\big( {V_{i_1}}, {V_{i_2}, V_{i_3}}, {V_{i_4}, V_{i_5}} \big) \sim \big( {U_{0}}, {U_{j_1}, U_{j_2}}, {U_{j_3}, U_{j_4}} \big)
\label{p5}
\end{eqnarray}
where $\mathcal{V}_{n_1} = \{V_{i_1}, V_{i_2}, V_{i_3}\}$, $\mathcal{V}_{n_2} = \{V_{i_1},V_{i_4},V_{i_5}\}$, $\mathcal{U}_{n_1} = \{U_{0}, U_{j_1}, U_{j_2}\}$, $\mathcal{U}_{n_2} = \{U_{0}, U_{j_3},U_{j_4}\}$.
Because $S$ is uniformly chosen from the set of all full rank matrices, we have
\begin{eqnarray}
(V_{i_1}, V_{i_2}, V_{i_3}, V_{i_4}, V_{i_5}) \sim (V_1, V_2, V_3, V_4, V_5)
\label{p2}
\end{eqnarray}
Next we note that there is a bijection between 
\begin{eqnarray}
(U_0, U_{j_1}, U_{j_2}, U_{j_3}, U_{j_4})&\leftrightarrow&(U_0, U_1, U_2, U_3, U_4)
\end{eqnarray}
This is because $(U_0, U_{j_1}, U_{j_2}, U_{j_3}, U_{j_4})$ always includes $U_0$, two terms out of $U_1, U_2, U_6, U_7$ and two terms out of $U_3, U_4, U_8, U_9$.  But from any two terms of $U_1, U_2, U_6, U_7$ there is a bijection to $U_1, U_2$, and from any two terms of $U_3,U_4,U_8, U_9$ there is a bijection to $U_3, U_4$. Now since $S' = (U_0; U_1; U_2; U_3; U_4; U_5)$ is picked uniformly from $\mathcal{S}$, conditioned on any feasible value of $U_5$, $(U_0, U_1, U_2, U_3, U_4)$ is uniformly distributed over all possible values that preserve full rank for $S'$. Since $(U_0, U_{j_1}, U_{j_2}, U_{j_3}, U_{j_4})$ spans the same space as $(U_0, U_1, U_2, U_3, U_4)$, they have the same set of feasible values. The bijection between them then means that $(U_0, U_{j_1}, U_{j_2}, U_{j_3}, U_{j_4})$ is also uniformly distributed over all possibilities that preserve full rank for $S'$, conditioned on any feasible $U_5$. That means
\begin{eqnarray}
(U_0, U_{j_1}, U_{j_2}, U_{j_3}, U_{j_4}) \sim (U_0, U_1, U_2, U_3, U_4) \label{p3}
\end{eqnarray}
Finally, we note that $S$ and $S'$ are identically distributed, so we have
\begin{eqnarray}
(V_1, V_2, V_3, V_4, V_5) \sim (U_0, U_1, U_2, U_3, U_4) 
\label{p4}
\end{eqnarray}
Combining (\ref{p2}), (\ref{p3}) and (\ref{p4}), we arrive at (\ref{p5}) and (\ref{space_privacy}).

\subsection{Rate achieved is $3/5$}
The rate achieved is $12/20 = 3/5$, because we download 20 symbols in total (5 from each server) and the desired message size is 12 symbols. 

\section{Optimality of Rate $3/5$}
We  presented a scheme that achieves the rate $3/5$ for the setting $(K, N, T, K_c) = (2, 4, 2, 2)$ with the MDS storage code $(x,y)\rightarrow(x,y,x+y,x+2y)$. But is the scheme optimal? i.e., is the rate $3/5$ the highest rate possible for this setting? To settle this question we need an upper bound. So far the best information theoretic upper bound that we are able to prove is $8/13$\footnote{{\color{black}Remarkably, $8/13$ can be shown to be the capacity if the colluding sets of servers are restricted to servers $\{1,2\},\{2,3\}, \{3,4\},\{4,1\}$ (see Section \ref{sec:restricted1}).}} (see Section \ref{sec:converse1}), which leaves the information theoretic capacity open for this setting. However, let us define the notion of ``linear capacity" as the highest rate that can be achieved by any (scalar or vector) linear PIR scheme. It turns out that we are  able to settle the linear capacity.
\begin{theorem}\label{thm:linear}
For the MDS-TPIR problem with $(K, N, T, K_c) = (2, 4, 2, 2)$ and the MDS storage code $(x,y)\rightarrow(x,y,x+y,x+2y)$, the linear capacity is $3/5$.
\end{theorem}
{\it Proof:} Since the achievability of $3/5$ has already been shown, we are left to prove the converse, i.e., the upper bound.

Let ${\bf a},{\bf b},{\bf c},{\bf d}\in\mathbb{F}_p^{L/2\times 1}$ be i.i.d. uniform $L/2\times 1$ vectors over $\mathbb{F}_p$. Without loss of generality, the MDS storage code for message $W_k$ is represented as follows. 
\begin{eqnarray}
W_1 = ({\bf a}; {\bf b}) && W_2 = ({\bf c}; {\bf d}) \label{eq:ww}
\end{eqnarray}
and the storage is specified as
\begin{eqnarray}
(W_{11},W_{12}, W_{13}, W_{14}) &=& ({\bf a}, {\bf b}, {\bf a +b}, {\bf a+2b}) \notag\\
(W_{21},W_{22}, W_{23}, W_{24}) &=& ({\bf c}, {\bf d}, {\bf c +d}, {\bf c+2d}) \label{eq:w1}
\end{eqnarray}
The scheme is linear so that the download from each server consists of linear combinations of the stored symbols of both messages. Furthermore, without loss of generality, we assume that the scheme is symmetric\footnote{Any scheme can be made symmetric, e.g., by repeating the original scheme for each of the $N{!}$ permutations of the servers to retrieve a correspondingly expanded message of length $L'=N{!} L$.} and  the download from each server is comprised of $d\leq L/2$ independent symbols from each message. Therefore, the downloads can be expressed as
\begin{eqnarray}
A_n^{[k]}&=&V_{1n}^{[k]} W_{1n}+V_{2n}^{[k]} W_{2n}, \forall n \in [1:4], k\in[1:2]\\
\mbox{rank}(V_{1n}^{[k]})&=&\mbox{rank}(V_{2n}^{[k]})=d \label{eq:rankd}
\end{eqnarray}
where $V_{in}^{[k]}$ are $D/4\times L/2$  matrices that may be chosen randomly by the user (functions of $\mathcal{F}$). Clearly we must have $4d\geq L$ otherwise the $L$ symbols of the desired message cannot be recovered.
Define $\epsilon\geq 0$ such that 
\begin{eqnarray}
4d& =& L(1+\epsilon)
\end{eqnarray}
Without loss of generality, let us assume henceforth that $W_2$ is the desired message. For the next set of arguments, we focus only on the downloads corresponding to $W_2$, i.e., set all $W_1$ symbols to $0$. Further, let us use the notation $\mathsf{V}$ to represent the row span of the matrix $V$. The symbols downloaded from Server $n$ along $\mathsf{V} \subset \mathsf{V}^{[2]}_{2n}$, are called redundant if they can be expressed as linear combinations of symbols downloaded from other servers, i.e., they contribute no new information.
\begin{eqnarray}
H(V W_{2n}|V^{[2]}_{2n_1} W_{2n_1}, V^{[2]}_{2n_2} W_{2 n_2}, V^{[2]}_{2n_3}W_{2n_3},\mathcal{F}, V) =0 \label{u_in}
\end{eqnarray}
where $n, n_1, n_2, n_3$ are distinct indices in $[1:4]$.  Note that we download no more than a total of $L(1+\epsilon)$ (possibly dependent) symbols of $W_2$ from all $4$ servers, from which we must be able to decode all $L$ independent symbols of $W_2$. Therefore, we cannot have more than $\epsilon L$ redundant symbols. Therefore, for any $V$ that satisfies (\ref{u_in}) we must have 
\begin{eqnarray}
\dim(\mathsf{V}) &\leq& \epsilon L \label{eq:rank}
\end{eqnarray}

Next, let us consider the pairwise overlap between $\mathsf{V}^{[2]}_{2i}$ and $\mathsf{V}^{[2]}_{2j}$, $i < j, i, j \in [1:4]$. By the symmetry of the scheme, there exist $V_{ij}$, $\forall i,j\in[1:4], i\neq j$, and $\alpha \geq 0$ such that
\begin{eqnarray}
\mathsf{V}_{ij} = \mathsf{V}^{[2]}_{2i} \cap \mathsf{V}^{[2]}_{2j},~~
\dim(\mathsf{V}_{ij}) = \alpha d \label{eq:dimd}
\end{eqnarray}

The following lemma formalizes the intuition that the overlaps $\alpha$ must be small enough to ensure that we have enough independent symbols to recover $W_2$.
\begin{lemma}
\begin{eqnarray}
3\alpha d &\leq& d + 2\epsilon L \\
\mbox{Equivalently, } \alpha &\leq& \frac{1}{3} + \frac{8}{3} \left( \frac{\epsilon}{1+\epsilon} \right)
\label{adl}
\end{eqnarray}
\end{lemma}
{\it Proof:} 
First, we show that 
\begin{eqnarray}
\dim(\mathsf{V}_{12} \cap \mathsf{V}_{13}) \leq \epsilon L \label{eq:v123}
\end{eqnarray}
For any vector $v \in \mathsf{V}_{12} \cap \mathsf{V}_{13}$ (note that $v$ belongs simultaneously to $\mathsf{V}_{21}^{[2]}, \mathsf{V}_{22}^{[2]}, \mathsf{V}_{23}^{[2]}$), the symbol $v W_{23}$ (downloaded from Server 3) is redundant because it is a linear combination of downloads from servers 1 and 2,
\begin{eqnarray}
v({\bf c+d}) &=& v{\bf c}+v{\bf d}\\
\therefore v W_{23} &=& v W_{21} + v W_{22} \\
&\Rightarrow& H(v W_{23}|V^{[2]}_{21} W_{21}, V^{[2]}_{22} W_{2 2},\mathcal{F},v) = 0 \label{eq:in1}
\end{eqnarray}
From (\ref{eq:in1}) and (\ref{eq:rank}), we have (\ref{eq:v123}).

Second, we show that 
\begin{eqnarray}
\dim\big((\mathsf{V}_{12} \cup \mathsf{V}_{13}\big) \cap \mathsf{V}_{14}) \leq \epsilon L \label{eq:v1234}
\end{eqnarray} 
Consider any vector $v \in \mathsf{V}_{12}$. Because $v$ belongs to both $\mathsf{V}_{21}^{[2]}$ and $\mathsf{V}_{22}^{[2]}$, we have downloaded $v W_{21}=v{\bf c}$ and $v W_{22}=v{\bf d}$ from servers 1 and 2. Similarly, for any vector $v' \in \mathsf{V}_{13}$, we have downloaded $v' W_{21}=v'{\bf c}$ and $v' W_{23} = v'({\bf c+d})=v' W_{21} + v' W_{22}$ (from servers 1 and 3), from which we can recover $v' W_{21}=v'{\bf c}$ and $v' W_{22}=v'{\bf d}$. Consider now any vector $v^* \in (\mathsf{V}_{12} \cup \mathsf{V}_{13}\big) \cap \mathsf{V}_{14}$. Suppose $v^* = h_1 v + h_2 v', v \in \mathsf{V}_{12}, v' \in \mathsf{V}_{13}$ for constants $h_1, h_2$. The symbol $v^* W_{24}=v^*({\bf c+2d})$ (downloaded from Server 4) is redundant because it is a linear combination of downloads from servers 1, 2 and 3,
\begin{eqnarray}
v^* W_{24} &=& (h_1 v + h_2 v')({\bf c+2d})\\
&=& h_1 v{\bf c} + 2h_1 v {\bf d} + h_2 v' {\bf c} + 2h_2 v' {\bf d}\\
&=& h_1 v W_{21} + 2h_1 v W_{22} + h_2 v' W_{21} + 2h_2 v' W_{22}\\
&\Rightarrow&~ H(v^* W_{24} |V^{[2]}_{21} W_{21}, V^{[2]}_{22} W_{2 2}, V^{[2]}_{23} W_{23},\mathcal{F}, v^*) = 0 \label{eq:in2}
\end{eqnarray}
\noindent From (\ref{eq:in2}) and (\ref{eq:rank}), we have (\ref{eq:v1234}). Next, consider $\dim(\mathsf{V}_{12} \cup \mathsf{V}_{13})$.
\begin{eqnarray}
&&\lefteqn{\dim(\mathsf{V}_{12} \cup \mathsf{V}_{13}) } \\
~~&=& \dim(\mathsf{V}_{12}) + \dim(\mathsf{V}_{13}) - \dim(\mathsf{V}_{12} \cap \mathsf{V}_{13}) \\
&\overset{}{\geq}& 2\alpha d - \epsilon L ~~\left(\mbox{\small from}~(\ref{eq:dimd})(\ref{eq:v123})\right) \label{eq:v123u}
\end{eqnarray}
Finally, consider $\dim(\mathsf{V}_{12} \cup \mathsf{V}_{13} \cup \mathsf{V}_{14})$. 
\begin{eqnarray}
d &=& \dim(\mathsf{V}_{21}^{[2]}) \geq \dim(\mathsf{V}_{12} \cup \mathsf{V}_{13} \cup \mathsf{V}_{14}) \\
&=&  \dim(\mathsf{V}_{12} \cup \mathsf{V}_{13}) + \dim(\mathsf{V}_{14}) -  \dim\big( (\mathsf{V}_{12} \cup \mathsf{V}_{13}) \cap \mathsf{V}_{14}\big)  \\
&\overset{}{\geq}& ~2\alpha d- \epsilon L + \alpha d - \epsilon L  ~~~\left(\mbox{\small from}~(\ref{eq:v123u})(\ref{eq:dimd})(\ref{eq:v1234})\right)
\\
&\Rightarrow&  3\alpha d \leq d + 2\epsilon L 
\end{eqnarray}
\hfill\QED

\noindent We now proceed to complete the converse.
\begin{eqnarray}
&& D +o(L)L \geq H(A_{1:4}^{[1]} | \mathcal{F}, \mathcal{G}) + o(L)L \\
&\overset{(\ref{corr})}{=}& H(A_{1:4}^{[1]}, W_1 | \mathcal{F}, \mathcal{G}) \\
&\overset{(\ref{indep})}{=}& H(W_1) + H(A_1^{[1]} | W_1, \mathcal{F}, \mathcal{G}) 
+ H(A_{2:4}^{[1]} | W_1, A_1^{[1]} \mathcal{F}, \mathcal{G})   \\
&\geq& H(W_1) + H(A_1^{[1]} | W_1, \mathcal{F}, \mathcal{G}) + H(A_{3:4}^{[1]} | W_1, W_{21}, A_1^{[1]}, \mathcal{F}, \mathcal{G})  \\
&\overset{(\ref{storage_size})(\ref{query_det})(\ref{answer_det})}{=}& H(W_1) + H(A_1^{[1]} | W_1, \mathcal{F}, \mathcal{G}) 
+ H(A_{3:4}^{[1]} | W_1, W_{21}, \mathcal{F}, \mathcal{G})  \\
&\overset{(\ref{storage_size})(\ref{same})}{=}& H(W_1) + H(A_1^{[2]} | W_1, \mathcal{F}, \mathcal{G}) 
+ H(A_{3:4}^{[2]} | W_1, W_{21}, \mathcal{F}, \mathcal{G})   \\
&\overset{(\ref{eq:ww})(\ref{eq:w1})}{=}&H({\bf a,b}) + H(V_{21}^{[2]}{\bf c}| \mathcal{F}) + H(V_{23}^{[2]}({\bf c+d}),V_{24}^{[2]}({\bf c+2d}) | {\bf c}, \mathcal{F})  \\
&=&H({\bf a,b}) + H(V_{21}^{[2]}{\bf c}| \mathcal{F}) 
+ H(V_{23}^{[2]}{\bf d},2V_{24}^{[2]}{\bf d} | \mathcal{F})   \\
&\overset{(\ref{h2})}{=}& L + \dim(\mathsf{V}_{21}^{[2]}) + \dim(\mathsf{V}_{23}^{[2]} \cup \mathsf{V}_{24}^{[2]}) \\ 
&\overset{(\ref{eq:rankd})(\ref{eq:dimd})}{=}& L + d + 2d - \alpha d  \\
&\overset{(\ref{adl})}{\geq}& L + \left(3- \frac{1}{3} - \frac{8}{3} \left( \frac{\epsilon}{1+\epsilon} \right)\right)\frac{(1+\epsilon)L}{4}  \\
&=&  5L/3 
\end{eqnarray}
Letting $L\rightarrow\infty$, we have $R = L/D \leq 3/5$.

\hfill\QED

\section{Capacity of a Class of MDS-TPIR Instances}
Building upon the insights from the achievable scheme and linear converse presented in the previous sections, we are able  to settle the information theoretic capacity of a non-trivial class of MDS-TPIR instances.
\begin{theorem}\label{thm:class}
For the class of MDS-TPIR instances with $(K,N, T, K_c)=(2,N,T,N-1)$, with arbitrary $T,N$, the capacity is $C = \frac{N^2 - N}{2N^2 - 3N + T}$.
\end{theorem}

The case $T=N$ is trivial because if all servers collude then the situation is equivalent to the single database scenario, i.e., it is optimal to download everything, and the capacity is $1/K=1/2$.  For the remaining cases,  $T<N$, and the proof of converse is presented in Section \ref{sec:converse2}. The proof of achievability for $T=2$ setting appears in Section \ref{sec:ach_proof2} where we present a scheme with zero error. The proof of achievability for $T>2$ settings appears in Section \ref{sec:ach_proof} where we present a scheme with vanishing probability of error. The remainder of this section presents two examples (one with $T=2$ and one with $T=3$) to illustrate the key ideas.

\subsection{Example: Capacity achieving scheme for $(K, N, T, K_c)=(2, 4, 2,3)$}
Let us present a scheme that achieves the rate $6/11$, which is the capacity for this setting according to Theorem \ref{thm:class}. As evident from the description below, the scheme builds upon the ideas that were introduced for  Theorem \ref{thm:disprove}. 

\subsubsection{Message and Storage Code}
Let each message be comprised of $L=N(N-1) = 12$ independent symbols from a sufficiently large finite field $\mathbb{F}_p$. Define ${\bf a}\in \mathbb{F}_{p}^{4 \times 1}$ as the vector $(a_1; a_2; a_3; a_4)$ comprised of i.i.d. uniform symbols $a_i \in \mathbb{F}_p$. Vectors ${\bf b, c, d, e, f}$ are defined similarly. Messages $W_1, W_2$ are defined in terms of these vectors as follows.
\begin{eqnarray}
W_1 = ({\bf a}; {\bf b}; {\bf c}) & W_2 = ({\bf d}; {\bf e}; {\bf f})
\end{eqnarray}
The $(N-1,N) = (3,4)$ MDS storage code is specified as follows. 
\begin{eqnarray}
(W_{11}, W_{12}, W_{13}, W_{14}) &=& ({\bf a}, {\bf b}, {\bf c}, {\bf a+b+c}) \\
(W_{21}, W_{22}, W_{23}, W_{24}) &=& ({\bf d}, {\bf e}, {\bf f}, {\bf d+e+f})
\end{eqnarray}
Note that each server stores $4$ symbols for each message and any three serves store just enough information to recover both messages (MDS property is satisfied).

\subsubsection{Construction of Queries}
 The query to each server consists of $6$ vectors, the first three for $W_1$ (denoted as $Q_n^{[k]}(W_1)$) and the last three for $W_2$ (denoted as $Q_n^{[k]}(W_2)$). The queries and downloads for $W_k, k \in [1:2]$ are described next. We denote $W_k = ({\bf x}; {\bf y}; {\bf z})$. When $k = 1$, $({\bf x}; {\bf y}; {\bf z}) = ({\bf a}; {\bf b}; {\bf c})$ and when $k = 2$, $({\bf x}; {\bf y}; {\bf z}) = ({\bf d}; {\bf e}; {\bf f})$.

Denote the set of all full rank $4 \times 4$ matrices over $\mathbb{F}_p$ as $\mathcal{S}_4$. The user privately chooses two matrices $S, S'$, independently and uniformly from $\mathcal{S}_4$. Label the rows of $S$ as $V_1, V_2, V_3, V_4$, and the rows of $S'$ as $\overline{U}_1, \overline{U}_2, U_1, U_2$. Define the following sets 
\begin{eqnarray}
\begin{array}{lllllr}%\\
\mathcal{V}_1&=&\{&V_2, &V_3, &V_4\},\\
\mathcal{V}_2&=&\{V_1,& &V_3, &V_4\},\\
\mathcal{V}_3&=&\{V_1,&V_2, & &V_4\},\\
\mathcal{V}_4&=&\{V_1,&V_2, &V_3 &\},
\end{array}&&
\begin{array}{lllll}
\mathcal{U}_1&=& \{\overline{U}_1, \overline{U}_2, {U}_{1} \} \\%= \{\overline{U}_1, U_1, U_2 \} \\
\mathcal{U}_2&=&\{\overline{U}_1, \overline{U}_2, {U}_{2} \} \\%= \{\overline{U}_1, U_3, U_1+U_2 \}\\
\mathcal{U}_3&=&\{\overline{U}_1, \overline{U}_2, {U}_{3} \} \\%= \{\overline{U}_1, U_1+U_3, U_2+U_3 \}\\
\mathcal{U}_4&=& \{\overline{U}_1, \overline{U}_2, {U}_{4} \} \\%= \{\overline{U}_1, U_1+U_2+U_3, U_1+2U_2+2U_3 \}
\end{array}
\end{eqnarray}
where $U_3, U_4$ are obtained as follows.
\begin{eqnarray}
U_3 &=& U_1 + U_2, \\
U_4 &=& U_1 + 2U_2
\end{eqnarray}
A preview of the scheme is as follows. For Server $n \in [1:4]$, the vectors in $\mathcal{V}_n$ are for the desired message and the vectors in $\mathcal{U}_n$ are for the undesired message. Since $K_c = N-1 = 3$, and each query vector $V_i$ is used no more than three times, all queries for the desired message will return  independent symbols for a total of $12$ desired symbols. For the undesired message, the same query vector $\overline{U}_1$ is used $4$ times such that only $3$ independent symbols are produced. Similarly the $4$ uses of $\overline{U}_2$ produce only $3$ independent symbols. Thus all queries for the undesired message will produce at most $6 + 4 = 10$ independent undesired symbols. The $12$ independent desired symbols and $10$ undesired symbols will be resolved from a total of $12 + 10 = 22$ downloaded symbols, to achieve the rate $12/22 = 6/11$. Privacy is ensured by the observation that any $\mathcal{V}_n, \mathcal{V}_n', n\neq n'$ have two elements in common and similarly any $\mathcal{U}_n, \mathcal{U}_n', n\neq n'$ have two elements in common. We now proceed to the details.

{When $W_k$ is desired}, we have $\forall n \in [1:4]$,
{%\setlength{\mathindent}{0.1cm}
\begin{align}
{\mbox{Server}~n}:&&
Q_n^{[k]}(W_k) &= \mathbb{\pi}_n (\mathcal{V}_n),& A_n^{[k]}(W_k) &= Q_n^{[k]}(W_k) W_{kn}.
\end{align}
}
\indent {\it Desired Symbols Are Independent:} From $A_{1:4}^{[k]}(W_k)$, the user can recover the $12$ symbols $V_2{\bf x}, V_3{\bf x}, V_4{\bf x}$, $V_1{\bf y}, V_3{\bf y}, V_4{\bf y}, V_1{\bf z}, V_2{\bf z}, V_4{\bf z}, V_1({\bf x}+{\bf y}+{\bf z}), V_2({\bf x}+{\bf y}+{\bf z}), V_3({\bf x}+{\bf y}+{\bf z})$ and therefore all $12$ symbols (${\bf x};{\bf y};{\bf z}$) of $W_k$, since $S = (V_1;V_2;V_3;V_4)$ has full rank.

{When $W_k$ is undesired}, we have $\forall n \in [1:4]$,
{%\setlength{\mathindent}{0cm}
\begin{align}
{\mbox{Server}~n}: &&Q_n^{[k^c]}(W_k) &= \mathbb{\pi}_n' (\mathcal{U}_n), &A_n^{[k^c]}(W_k)& = Q_n^{[k^c]}(W_k) W_{kn}.
\end{align}
\indent{\it Interfering Symbols Are Dependent and Have Dimension at most $10$:}
Consider the interfering symbols along the common vectors $\overline{U}_1, \overline{U}_2$. Note that 
\begin{eqnarray}
%\hspace{1cm}
\overline{U}_1 {\bf x} + \overline{U}_1 {\bf y} + \overline{U}_1 {\bf z} = \overline{U}_1 ({\bf x} + {\bf y} + {\bf z})\\
\overline{U}_2 {\bf x} + \overline{U}_2 {\bf y} + \overline{U}_2 {\bf z} = \overline{U}_2 ({\bf x} + {\bf y} + {\bf z})
\end{eqnarray}
Since at least $2$ interfering symbols are linear combinations of the rest, the $12$ interfering symbols cannot have more than $10$ dimensions, i.e., their joint entropy is no more than $10$ in $p$-ary units. 

\subsubsection{Combining Answers, Correctness and Rate}

The combining process and correctness proof are similar to that in Theorem \ref{thm:disprove}. The difference is that in Theorem \ref{thm:disprove}, we find the explicit choice of combining matrices, here we will only prove the existence of combining matrices over a sufficiently large field. The details are deferred to the general proof in Section \ref{sec:ach_proof2}. We repeat the above query construction two times independently such that each server has $6 \times 2 = 12$ symbols ($6$ in $W_1$ and $6$ in $W_2$).
These $12$ symbols at each server are combined to $11$ downloaded symbols, $A_n^{[k]}$ and it is ensured that we can decode all interfering symbols and then extract the desired symbols. 

Thus, the rate achieved is $6/11$. 

\subsubsection{Privacy Proof}
The privacy proof is virtually identical to that in Theorem \ref{thm:disprove}, so the details are deferred to the general proof in Section \ref{sec:ach_proof2}.

{\subsection{Example: Capacity achieving scheme for $(K, N, T, K_c)=(2, 4,3,3)$}
Let us present a scheme that achieves the rate $12/23$, which is the capacity for this setting according to Theorem \ref{thm:class}.  The key distinction of this $T = 3$ case with the $T=2$ case presented in the previous section is that permutations of the query vectors are no longer enough to ensure the privacy. So we will resort to sending the space spanned by the query vectors instead of the query vectors themselves. Furthermore, instead of guaranteeing zero-error, we will only show that the probability of error can be made arbitrarily small by choosing a sufficiently large message size.

\subsubsection{Message and Storage Code}
The message construction and storage code are the same as  when $T=2$.
Let each message be comprised of $L=N(N-1) = 12$ independent symbols from a sufficiently large finite field $\mathbb{F}_p$. Define ${\bf a}\in \mathbb{F}_{p}^{4 \times 1}$ as the vector $(a_1; a_2; a_3; a_4)$ comprised of i.i.d. uniform symbols $a_i \in \mathbb{F}_p$. Vectors ${\bf b, c, d, e, f}$ are defined similarly. Messages $W_1, W_2$ are defined in terms of these vectors as follows.
\begin{eqnarray}
W_1 = ({\bf a}; {\bf b}; {\bf c}) & W_2 = ({\bf d}; {\bf e}; {\bf f})
\end{eqnarray}
The $(N-1,N) = (3,4)$ MDS storage code is specified as follows. 
\begin{eqnarray}
(W_{11}, W_{12}, W_{13}, W_{14}) &=& ({\bf a}, {\bf b}, {\bf c}, {\bf a+b+c}) \\
(W_{21}, W_{22}, W_{23}, W_{24}) &=& ({\bf d}, {\bf e}, {\bf f}, {\bf d+e+f})
\end{eqnarray}

\subsubsection{Construction of Queries}
The query to each server consists of two vector spaces, one for $W_1$ (span of the rows of $Q_n^{[k]}(W_1)$) and one for $W_2$ (span of the rows of  $Q_n^{[k]}(W_2)$). The queries and downloads for $W_k, k \in [1:2]$ are described next. We denote $W_k = ({\bf x}; {\bf y}; {\bf z})$. When $k = 1$, $({\bf x}; {\bf y}; {\bf z}) = ({\bf a}; {\bf b}; {\bf c})$ and when $k = 2$, $({\bf x}; {\bf y}; {\bf z}) = ({\bf d}; {\bf e}; {\bf f})$.

Denote the set of all full rank $4 \times 4$ matrices over $\mathbb{F}_p$ as $\mathcal{S}_4$. The user privately chooses two matrices $S, S'$, independently and uniformly from $\mathcal{S}_4$. Label the rows of $S$ as $V_1, V_2, V_3, V_4$, and the rows of $S'$ as $\overline{U}_1, U_1, U_2, U_3$. Define the following sets \begin{eqnarray}
\begin{array}{lllllr}%\\
\mathcal{V}_1&=&\{&V_2, &V_3, &V_4\},\\
\mathcal{V}_2&=&\{V_1,& &V_3, &V_4\},\\
\mathcal{V}_3&=&\{V_1,&V_2, & &V_4\},\\
\mathcal{V}_4&=&\{V_1,&V_2, &V_3 &\},
\end{array}&&
\begin{array}{lllll}
\mathcal{U}_1&=& \{\overline{U}_1, \widetilde{U}_{1}, \widetilde{U}_{2} \} = \{\overline{U}_1, U_1, U_2 \} \\
\mathcal{U}_2&=&\{\overline{U}_1, \widetilde{U}_{3}, \widetilde{U}_{4} \} = \{\overline{U}_1, U_3, U_1+U_2 \}\\
\mathcal{U}_3&=&\{\overline{U}_1, \widetilde{U}_{5}, \widetilde{U}_{6} \} = \{\overline{U}_1, U_1+U_3, U_2+U_3 \}\\
\mathcal{U}_4&=& \{\overline{U}_1, \widetilde{U}_{7}, \widetilde{U}_{8} \} = \{\overline{U}_1, U_1+U_2+U_3, U_1+2U_2+2U_3 \}
\end{array}
\end{eqnarray}
%
%\begin{eqnarray}
%\mathcal{V}_1 = \{V_2, V_3, V_4\}, && \mathcal{U}_1 = \{\overline{U}_1, \widetilde{U}_{1}, \widetilde{U}_{2} \} = \{\overline{U}_1, U_1, U_2 \} \\
%\mathcal{V}_2 = \{V_1, V_3, V_4\}, && \mathcal{U}_2 = \{\overline{U}_1, \widetilde{U}_{3}, \widetilde{U}_{4} \} = \{\overline{U}_1, U_3, U_1+U_2 \}\\
%\mathcal{V}_3 = \{V_1, V_2, V_4\}, && \mathcal{U}_3 = \{\overline{U}_1, \widetilde{U}_{5}, \widetilde{U}_{6} \} = \{\overline{U}_1, U_1+U_3, U_2+U_3 \}\\
%\mathcal{V}_4 = \{V_1, V_2, V_3\}, && \mathcal{U}_4 = \{\overline{U}_1, \widetilde{U}_{7}, \widetilde{U}_{8} \} = \{\overline{U}_1, U_1+U_2+U_3, U_1+2U_2+2U_3 \}
%\end{eqnarray}
where $\widetilde{U}_1, \cdots, \widetilde{U}_{8}$ are the rows of $\widetilde{U}$, obtained as follows.
\begin{eqnarray}
\widetilde{U} &=& {P} (U_1; U_2; U_3) \label{uu}\\
\mbox{i.e.,}\left(\begin{array}{c}
\widetilde{U}_{1}\\
\widetilde{U}_{2}\\
\widetilde{U}_{3}\\
\widetilde{U}_{4}\\
\widetilde{U}_{5}\\
\widetilde{U}_{6}\\
\widetilde{U}_{7}\\
\widetilde{U}_{8}
\end{array}\right) &=& \left(
\begin{array}{ccc}
1 & 0 & 0 \\
0 & 1 & 0 \\
0 & 0 & 1 \\
1 & 1 & 0 \\
1 & 0 & 1 \\
0 & 1 & 1 \\
1 & 1 & 1 \\
1 & 2 & 2 
\end{array}
\right)\left(\begin{array}{c}U_1\\U_2\\U_3\end{array}\right)
\end{eqnarray}
A preview of the scheme is as follows. For Server $n \in [1:4]$, the span of $\mathcal{V}_n$ is the query space for the desired message and the span of $\mathcal{U}_n$ is the query space for the undesired message. Since $K_c = N-1 = 3$, and each query vector $V_i$ is used no more than three times, all queries for the desired message will return  independent symbols for a total of $12$ desired symbols. For the undesired message, the same query vector $\overline{U}_1$ is used $4$ times such that only $3$ independent symbols will be produced. Thus all queries for the undesired message will produce at most $3 + 8 = 11$ independent undesired symbols. The $12$ independent desired symbols and $11$ undesired symbols will be resolved from a total of $12 + 11 = 23$ downloaded symbols, to achieve the rate 12/23. Privacy is ensured by choosing ${P}$ in such a way that it allows a bijective mapping between the $\mathcal{U}_n$ or $\mathcal{V}_n$ spaces that may be observed by any set of up to $T = 3$ colluding servers. The bijection shows that the queries for both desired and undesired messages are uniformly distributed, and therefore indistinguishable. While a specific $P$ is chosen for this example, there are many choices of $P$ that will work. In fact, $P$ only needs to be sufficiently generic, so as the field size grows, almost all choices of $P$ will work. We now proceed to the details.

{When $W_k$ is desired}, we have $\forall n \in [1:4]$,
{%\setlength{\mathindent}{0.1cm}
\begin{align}
{\mbox{Server}~n}:&&
Q_n^{[k]}(W_k) &= \mathbb{B} (\mathcal{V}_n),& A_n^{[k]}(W_k) &= Q_n^{[k]}(W_k) W_{kn}.
\end{align}
}where $\mathbb{B}(\mathcal{V})$ represents the reduced row echelon form of a matrix whose rows are the elements of  $\mathcal{V}$. The reduced row echelon form ensures that the queries reveal only the space spanned by the corresponding $V_i$ vectors to each server, and not  directly the $V_i$ vectors themselves.

\indent {\it Desired Symbols Are Independent:} From $A_{1:4}^{[{k}]}(W_k)$, we can recover the $12$ symbols of $W_k$. 
Note that because the user knows $V_{1:4}$, from $A_{1:4}^{[k]}(W_k)$ he can recover the projections along $V_i$. For example, the row reduced echelon form for $\mathcal{V}_1$ is a change of basis operation that can be represented as $\mathbb{B}(\mathcal{V}_1)=B_1(V_2;V_3;V_4)$ for some invertible matrix $B_1$. Since the user knows $B_1$, he can multiply $A_{1}^{[{k}]}(W_k)$ with $(B_1)^{-1}$ as follows
\begin{eqnarray}
B_1^{-1}A_{1}^{[{k}]}(W_k)&=&B_1^{-1}Q_n^{[k]}(W_k) W_{k1}\\
&=&B_1^{-1}B_1(V_2;V_3;V_4){\bf x}\\
&=&(V_2{\bf x}; V_3{\bf x}; V_4{\bf x})
\end{eqnarray}
Thus, from $A_{1:4}^{[k]}(W_k)$ the user recovers the $12$ symbols $V_2{\bf x}, V_3{\bf x}, V_4{\bf x}$, $V_1{\bf y}, V_3{\bf y}, V_4{\bf y}, V_1{\bf z}, V_2{\bf z}, V_4{\bf z}, V_1({\bf x}+{\bf y}+{\bf z}), V_2({\bf x}+{\bf y}+{\bf z}), V_3({\bf x}+{\bf y}+{\bf z})$ and therefore all $12$ symbols (${\bf x};{\bf y};{\bf z}$) of $W_k$, since $S = (V_1;V_2;V_3;V_4)$ has full rank.

{When $W_k$ is undesired}, we have $\forall n \in [1:4]$,
{%\setlength{\mathindent}{0cm}
\begin{align}
{\mbox{Server}~n}: &&Q_n^{[k^c]}(W_k) &= \mathbb{B} (\mathcal{U}_n), &A_n^{[k^c]}(W_k)& = Q_n^{[k^c]}(W_k) W_{kn}.
\end{align}
\indent{\it Interfering Symbols Are Dependent and Have Dimension at most $11$:}
Consider the interfering symbols along the common vector $\overline{U}_1$. Note that 
\begin{eqnarray}
%\hspace{1cm}
\overline{U}_1 {\bf x} + \overline{U}_1 {\bf y} + \overline{U}_1 {\bf z} = \overline{U}_1 ({\bf x} + {\bf y} + {\bf z})
\end{eqnarray}
Since at least $1$ interfering symbol is a linear combination of the rest, the $12$ interfering symbols cannot have more than $11$ dimensions, i.e., their joint entropy is no more than $11$ in $p$-ary units. 

\subsubsection{Combining Answers, Correctness and Rate}
The combining process and correctness proof are similar to that in Theorem 1 except that the combining matrices $C_n$ are chosen in a uniformly random manner now (so the matrices are no longer deterministic). We will show in Section \ref{sec:ach_proof} that independent and uniformly random choices of $C_n$ are enough to guarantee that as the field size approaches infinity, i.e., $p\rightarrow\infty$, the probability of error, $P_e\rightarrow 0$. The reasoning for the rate calculation is as follows. We repeat the above query construction four times independently such that each server has $6 \times 4 = 24$ symbols ($12$ in $W_1$ and $12$ in $W_2$). These $24$ symbols at each server are combined to $23$ downloaded symbols, $A_n^{[k]}$ and it is ensured that we can almost surely decode all interfering symbols and then extract the desired symbols.  Thus, the rate achieved is $12/23$. 

\subsubsection{Privacy Proof}
Since the privacy proof is a bit more involved now, let us use this example to introduce the key ideas.
To show that the scheme is private to any $T = 3$ colluding servers, it suffices to show that the queries for $W_k$ for any $T = 3$ servers are identically distributed, regardless of which message is desired. Consider 3 distinct indices $i , j , l, i < j <l$ in $[1:4]$, we require 
\begin{eqnarray}
\left(Q_{i}^{[k]}(W_k), Q_{j}^{[k]}(W_k), Q_{l}^{[k]}(W_k) \right) &\sim& \left(Q_{i}^{[k^c]}(W_k), Q_{j}^{[k^c]}(W_k), Q_{l}^{[k^c]}(W_k) \right) \\
\Longleftrightarrow ~~~~ \left(\mathbb{B}(\mathcal{V}_i), \mathbb{B}(\mathcal{V}_j), \mathbb{B}(\mathcal{V}_l) \right) &\sim& \left( \mathbb{B}(\mathcal{U}_i), \mathbb{B}(\mathcal{U}_j), \mathbb{B}(\mathcal{U}_l)  \right) \label{space_privacy_ex}
\end{eqnarray}
Note that 
\begin{eqnarray}
\left(\mathbb{B}(\mathcal{V}_i), \mathbb{B}(\mathcal{V}_j), \mathbb{B}(\mathcal{V}_l) \right) = \left(\mathbb{B}(\{V_m, V_j, V_l\}), \mathbb{B}(\{V_m, V_i, V_l\}), \mathbb{B}(\{V_m, V_i, V_j\}) \right) \label{eq:vijl}
\end{eqnarray}
where $m \notin \{i,j,l\}, m \in [1:4]$. 
To prove (\ref{space_privacy_ex}), we wish to transform the spaces on the RHS to the form that is the same as (\ref{eq:vijl}). %the spaces on the LHS. 
To this end, we first compute the vectors that lie in the span of both $\mathbb{B}(\mathcal{U}_i)$ and $\mathbb{B}(\mathcal{U}_j)$, $i < j$. Note that the matrix $P$ is designed such that except $\overline{U}_{1}$, we have only one such vector (up to scaling), denoted as $U_{\{i,j\}}$. $U_{\{i,j\}}$ are computed explicitly as follows. Further, we fix the scaling factor such that the $U_{\{i,j\}}$ vector is unique.
\begin{eqnarray}
U_{\{1,2\}} &=& U_1 + U_2 \\
U_{\{1,3\}} &=& U_1 - U_2 \\
U_{\{1,4\}} &=& U_1 \\
U_{\{2,3\}} &=& U_1 + U_2 + 2U_3 \\
U_{\{2,4\}} &=& U_1 + U_2 + U_3 \\
U_{\{3,4\}} &=& U_2 + U_3 
\end{eqnarray}
It is easy to verify that $U_{\{i,j\}}, U_{\{i,l\}}, U_{\{j,l\}}$, $i,j,l \in [1:4], i < j < l$, are linearly independent, i.e.,
\begin{eqnarray}
(i,j,l) = (1,2,3) && \mbox{rank}(U_{\{1,2\}}; U_{\{1,3\}} ; U_{\{2,3\}}) = \mbox{rank}(U_1 + U_2; U_1 - U_2; U_1 + U_2 + 2U_3) = 3 \notag \\
(i,j,l) = (1,2,4) && \mbox{rank}(U_{\{1,2\}}; U_{\{1,4\}}; U_{\{2,4\}}) = \mbox{rank}(U_1 + U_2; U_1; U_1 + U_2 + U_3) = 3 \notag \\
(i,j,l) = (1,3,4) && \mbox{rank}(U_{\{1,3\}}; U_{\{1,4\}}; U_{\{3,4\}}) = \mbox{rank}(U_1 - U_2; U_1; U_2 + U_3) = 3 \notag \\
(i,j,l) = (2,3,4) && \mbox{rank}(U_{\{2,3\}}; U_{\{2,4\}}; U_{\{3,4\}}) = \mbox{rank}(U_1 + U_2 + 2U_3; U_1 + U_2 + U_3; U_2 + U_3) = 3 \notag \\
\label{eq:u123}
\end{eqnarray}
As a result, we may equivalently represent $Q_i^{[k^c]}(W_k)$ as
\begin{eqnarray}
Q_i^{[k^c]}(W_k) =  \mathbb{B}(\mathcal{U}_i) = \mathbb{B}(\{\overline{U}_1, U_{\{i,j\}}, U_{\{i,l\}}\}), \forall  i,j,l \in [1:4], i \neq j, i\neq l, j \neq l \label{eq:rep}
\end{eqnarray}
Note that equipped with this representation, $\left( \mathbb{B}(\mathcal{U}_i), \mathbb{B}(\mathcal{U}_j), \mathbb{B}(\mathcal{U}_l)  \right)$ is now of the same form as $\left(\mathbb{B}(\mathcal{V}_i), \mathbb{B}(\mathcal{V}_j), \mathbb{B}(\mathcal{V}_l) \right)$   and we are now ready to prove the privacy condition (\ref{space_privacy_ex}). %Using the representation in (\ref{eq:rep}), we have
\begin{eqnarray}
(\ref{space_privacy_ex}) \Longleftrightarrow && \left(\mathbb{B}(\{V_m, V_j, V_l\}), \mathbb{B}(\{V_m, V_i, V_l\}), \mathbb{B}(\{V_m, V_i, V_j\}) \right) \notag \\
&\sim& \left( \mathbb{B}(\{\overline{U}_1, U_{\{i,j\}}, U_{\{i,l\}}\}), \mathbb{B}(\{\overline{U}_1, U_{\{i,j\}}, U_{\{j,l\}}\}), \mathbb{B}(\{\overline{U}_1, U_{\{i,l\}}, U_{\{j,l\}}\})  \right)
\end{eqnarray}
Therefore, it suffices to show the following.
\begin{eqnarray}
(V_m, V_i, V_j, V_l) \sim (\overline{U}_1, U_{\{j,l\}}, U_{\{i,l\}}, U_{\{i,j\}}) \label{eq:pp4}
\end{eqnarray}
Because $S$ is uniformly chosen from the set of all full rank matrices, we have
\begin{eqnarray}
(V_m, V_i, V_j, V_l) \sim (V_1, V_2, V_3, V_4) \label{eq:pp3}
\end{eqnarray}
Based on  (\ref{eq:u123}),  there is a bijection between 
\begin{eqnarray}
(\overline{U}_1, U_{\{j,l\}}, U_{\{i,l\}}, U_{\{i,j\}}) \leftrightarrow  (\overline{U}_1, U_1, U_2, U_3)
\end{eqnarray}
Now since $S' = (\overline{U}_1; U_1; U_2; U_3)$ is uniform in all full rank matrices, the above bijection then means that $(\overline{U}_1; U_{\{j,l\}}; U_{\{i,l\}}; U_{\{i,j\}})$ is also uniform in all full rank matrices, i.e.,
\begin{eqnarray}
(\overline{U}_1, U_{\{j,l\}}, U_{\{i,l\}}, U_{\{i,j\}}) \sim  (\overline{U}_1, U_1, U_2, U_3) \label{eq:pp2}
\end{eqnarray}
Finally, note that $S$ and $S'$ have the same distribution, so we have
\begin{eqnarray}
(V_1, V_2, V_3, V_4) \sim  (\overline{U}_1, U_1, U_2, U_3) \label{eq:pp1}
\end{eqnarray}
Therefore, from (\ref{eq:pp3}), (\ref{eq:pp2}) and (\ref{eq:pp1}), we have proved (\ref{eq:pp4}) and (\ref{space_privacy_ex}).
%Note that the above example (in particular, the specification of $P$) works over the field $\mathbb{F}_3$. 
}}
\hfill\QED

\section{Conclusion} \label{sec:disc}
We settle a conjecture on the capacity of MDS-TPIR by Freij-Hollanti et al.  \cite{FREIJ_HOLLANTI} by constructing a scheme that  beats the conjectured capacity for one particular instance of MDS-TPIR. The rate achieved by the new scheme is shown to be the best possible rate that can be achieved by any linear scheme for the same MDS storage code. The insights from the achievability and converse arguments allow us to characterize the capacity of a class of MDS-TPIR instances. Through another counterexample, we are also able to prove that the  capacity expression cannot be symmetric in $T$ and $K_c$ parameters, i.e., these parameters cannot be interchangeable in general. Nevertheless, the general capacity expression for MDS-TPIR remains unknown.

\section{Appendix}
\subsection{Examples of Optimal Schemes over Small Fields}\label{sec:ex_small_field}
To highlight that the assumption of large field size (which was made convenience) may not be essential, in this section, we provide two examples of explicit MDS-TPIR capacity achieving schemes over small fields. 

\subsubsection{\bf Example 1}
Consider the MDS-TPIR instance with $(K,N,T,K_c) = (2,3,2,2)$. Note that the capacity of this setting is $6/11$, as established in Theorem \ref{thm:class}. We provide an alternative achievable scheme for rate $6/11$. In particular, the scheme operates over the binary field and the upload is 4 bits per server (the query to each server takes values in a set with cardinality $2^4 = 16$). %and the MDS storage code is a vector (block) code in that .

We assume that each message is $L = 6$ bits. Denote $a_1, \cdots, a_6, b_1, \cdots, b_6$ as 12 i.i.d. uniform bits, $a_i, b_i \in \mathbb{F}_2$. Messages $W_1, W_2$ are defined in terms of these bits as follows.
{%\setlength{\mathindent}{0cm}
\begin{eqnarray}
W_1 = (a_1; a_2; a_3; a_4; a_5; a_6), W_2 = (b_1; b_2; b_3; b_4; b_5; b_6)
\end{eqnarray}}
The storage is specified as
\begin{eqnarray}
{\mbox{Server 1}}:&&W_{11} = (a_1; a_2; a_3), W_{21} = (b_1; b_2; b_3) \\
{\mbox{Server 2}}:&&W_{12} = (a_4; a_5; a_6), W_{22} = (b_4; b_5; b_6) \\
{\mbox{Server 3}}:&&W_{13} = (\alpha_1; \alpha_2; \alpha_3), W_{23} = (\beta_1; \beta_2; \beta_3)
\end{eqnarray}
where $\alpha_1, \alpha_2, \alpha_3, \beta_1, \beta_2, \beta_3$ are obtained as follows.
\begin{eqnarray}
\alpha_1 = a_1 + a_2 + a_5, &&\beta_1 = b_1 + b_2 + b_5 \\
\alpha_2 = a_1 + a_3 + a_6, &&\beta_2 = b_1 + b_3 + b_6 \\
\alpha_3 = a_2 + a_4 + a_6, &&\beta_3 = b_2 + b_4 + b_6 
\end{eqnarray}
Further define
\begin{eqnarray}
\alpha_4 &=& \alpha_1 + \alpha_2 + \alpha_3 = a_3 + a_4 + a_5 \\
\beta_4 &=& \beta_1 + \beta_2 + \beta_3 = b_3 + b_4 + b_5
\end{eqnarray}
Note that each server stores 3 bits of each message and the storage at any 2 servers is just enough to recover both messages (MDS storage property is satisfied).

Define a function that maps 4 input bits to 3 output bits as follows.
\begin{eqnarray}
\mathcal{L}_3 (X_1, X_2, Y_1, Y_2) = (X_1+Y_2, X_2+Y_2, Y_1+Y_2)
\end{eqnarray}

We now describe the PIR scheme. $\mathcal{F}$ is a uniform random variable in $[1:16]$. Depending on the value of $\mathcal{F}$ and the desired message index $\theta \in [1:2]$, the user's query is specified by Table 1. The double-quotes notation around a random variable represents the \emph{query} about its realization. Note that the queries to Server $1$ and Server $2$ are the same, regardless of the value of $\theta$ and the query to Server $3$ is a deterministic function of that to Server $1$ and Server $2$.

\begin{table}[h]
\centering
\caption{\small The Scheme for MDS-TPIR with $(K,N,T,K_c) = (2,3,2,2)$.}
\begin{tabular}{|c|c|c|c|c|c|c|}\hline
$\mathcal{F}$&\mbox{Prob.}&$Q_1^{[\theta]}$~(\mbox{Server $1$})&$Q_2^{[\theta]}$~(\mbox{Server $2$})&$Q_3^{[1]}$~(\mbox{Server 3})&$Q_3^{[2]}$~(\mbox{Server 3})\\\hline
1&$1/16$&``$a_1,a_2,b_1,b_2$''&``$a_4,a_5,b_4,b_5$''&``$\mathcal{L}_3(\alpha_3, \alpha_4,\beta_1, \beta_2)$''&``$\mathcal{L}_3(\alpha_1, \alpha_2,\beta_3, \beta_4)$''\\
\cline{3-6}
2&$1/16$&``$a_1,a_3,b_1,b_2$''&``$a_4,a_5,b_4,b_5$''&``$\mathcal{L}_3(\alpha_1, \alpha_2,\beta_1, \beta_2)$''&``$\mathcal{L}_3(\alpha_3, \alpha_4,\beta_3, \beta_4)$''\\\cline{3-6}
3&$1/16$&``$a_1,a_2,b_1,b_3$''&``$a_4,a_5,b_4,b_5$''&``$\mathcal{L}_3(\alpha_3, \alpha_4,\beta_3, \beta_4)$''&``$\mathcal{L}_3(\alpha_1, \alpha_2,\beta_1, \beta_2)$''\\\cline{3-6}
4&$1/16$&``$a_1,a_3,b_1,b_3$''&``$a_4,a_5,b_4,b_5$''&``$\mathcal{L}_3(\alpha_1, \alpha_2,\beta_3, \beta_4)$''&``$\mathcal{L}_3(\alpha_3, \alpha_4,\beta_1, \beta_2)$''\\\cline{3-6}\\
\cline{3-6}
5&$1/16$&``$a_1,a_2,b_1,b_2$''&``$a_4,a_6,b_4,b_5$''&``$\mathcal{L}_3(\alpha_1, \alpha_2,\beta_1, \beta_2)$''&``$\mathcal{L}_3(\alpha_3, \alpha_4,\beta_3, \beta_4)$''\\\cline{3-6}
6&$1/16$&``$a_1,a_3,b_1,b_2$''&``$a_4,a_6,b_4,b_5$''&``$\mathcal{L}_3(\alpha_3, \alpha_4,\beta_1, \beta_2)$''&``$\mathcal{L}_3(\alpha_1, \alpha_2,\beta_3, \beta_4)$''\\\cline{3-6}
7&$1/16$&``$a_1,a_2,b_1,b_3$''&``$a_4,a_6,b_4,b_5$''&``$\mathcal{L}_3(\alpha_1, \alpha_2,\beta_3, \beta_4)$''&``$\mathcal{L}_3(\alpha_3, \alpha_4,\beta_1, \beta_2)$''\\\cline{3-6}
8&$1/16$&``$a_1,a_3,b_1,b_3$''&``$a_4,a_6,b_4,b_5$''&``$\mathcal{L}_3(\alpha_3, \alpha_4,\beta_3, \beta_4)$''&``$\mathcal{L}_3(\alpha_1, \alpha_2,\beta_1, \beta_2)$''\\\cline{3-6}\\
\cline{3-6}
9&$1/16$&``$a_1,a_2,b_1,b_2$''&``$a_4,a_5,b_4,b_6$''&``$\mathcal{L}_3(\alpha_3, \alpha_4,\beta_3, \beta_4)$''&``$\mathcal{L}_3(\alpha_1, \alpha_2,\beta_1, \beta_2)$''\\\cline{3-6}
10&$1/16$&``$a_1,a_3,b_1,b_2$''&``$a_4,a_5,b_4,b_6$''&``$\mathcal{L}_3(\alpha_1, \alpha_2,\beta_3, \beta_4)$''&``$\mathcal{L}_3(\alpha_3, \alpha_4,\beta_1, \beta_2)$''\\\cline{3-6}
11&$1/16$&``$a_1,a_2,b_1,b_3$''&``$a_4,a_5,b_4,b_6$''&``$\mathcal{L}_3(\alpha_3, \alpha_4,\beta_1, \beta_2)$''&``$\mathcal{L}_3(\alpha_1, \alpha_2,\beta_3, \beta_4)$''\\\cline{3-6}
12&$1/16$&``$a_1,a_3,b_1,b_3$''&``$a_4,a_5,b_4,b_6$''&``$\mathcal{L}_3(\alpha_1, \alpha_2,\beta_1, \beta_2)$''&``$\mathcal{L}_3(\alpha_3, \alpha_4,\beta_3, \beta_4)$''\\\cline{3-6}\\
\cline{3-6}
13&$1/16$&``$a_1,a_2,b_1,b_2$''&``$a_4,a_6,b_4,b_6$''&``$\mathcal{L}_3(\alpha_1, \alpha_2,\beta_3, \beta_4)$''&``$\mathcal{L}_3(\alpha_3, \alpha_4,\beta_1, \beta_2)$''\\\cline{3-6}
14&$1/16$&``$a_1,a_3,b_1,b_2$''&``$a_4,a_6,b_4,b_6$''&``$\mathcal{L}_3(\alpha_3, \alpha_4,\beta_3, \beta_4)$''&``$\mathcal{L}_3(\alpha_1, \alpha_2,\beta_1, \beta_2)$''\\\cline{3-6}
15&$1/16$&``$a_1,a_2,b_1,b_3$''&``$a_4,a_6,b_4,b_6$''&``$\mathcal{L}_3(\alpha_1, \alpha_2,\beta_1, \beta_2)$''&``$\mathcal{L}_3(\alpha_3, \alpha_4,\beta_3, \beta_4)$''\\\cline{3-6}
16&$1/16$&``$a_1,a_3,b_1,b_3$''&``$a_4,a_6,b_4,b_6$''&``$\mathcal{L}_3(\alpha_3, \alpha_4,\beta_1, \beta_2)$''&``$\mathcal{L}_3(\alpha_1, \alpha_2,\beta_3, \beta_4)$''\\\hline
\end{tabular}
\end{table}

We show that the scheme is both correct and private. The schemes is correct because our scheme satisfies the important property (${\it P1}$) that from the answers $A_1^{[k]}, A_2^{[k]}$, we always know one undesired bit in $A_3^{[k]}$ and then we can extract the $2$ desired bits in $A_3^{[k]}$ (because if any $1$ of the $4$ input bits of the $\mathcal{L}_3$ function is known, the remaining $3$ input bits can be solved from the $3$ output bits). Combining these $2$ desired bits with the other $4$ desired bits (2 from Server $1$ and $2$ from Server $2$), we obtain the desired message (easy to verify that these 6 bits are independent).
The property ($\it P1$) is easy to verify. For example, consider $k=1$ and $\mathcal{F} = 8$. From $A_1^{[1]}, A_2^{[1]}$, we obtain $b_1, b_3, b_4, b_5$, from which we further obtain $\beta_4 = b_3 + b_4 + b_5$ and $\beta_4$ appears in $A_3^{[1]}$. The scheme is private because it is easy to verify that for any $2$ servers, the queries are identically distributed no matter which message is desired and then the privacy condition (\ref{privacy}) is satisfied.

The scheme downloads $4$ bits from Server $1$, $4$ bits from Server $2$ and $3$ bits from Serve $3$. It retrieves $6$ desired message bits. Therefore the rate is $6/11$.

\subsubsection{\bf Example 2} Consider the MDS-TPIR instance with $(K,N,T,K_c) = (2,4,3,2)$.\label{sec:Ex2}
The capacity of this setting turns out to be $4/7$. The rate can not be more than $4/7$ because the capacity of TPIR with $(K,N,T) = (2,4,3)$ is $4/7$ \cite{Sun_Jafar_TPIR} and reducing $K_c$ from $2$ to $1$ can not hurt. We provide an achievable scheme for rate $4/7$. In particular, the scheme operates over the finite field $\mathbb{F}_{13}$ and the upload is 6 bits per server (the query to each server takes values in a set with cardinality $2^6 = 64$). %and the MDS storage code is a vector (block) code in that .

We assume that each message is $L = 4$ symbols. Denote $a_1, a_2, a_3, a_4, b_1, b_2, b_3, b_4$ as 8 i.i.d. uniform symbols, $a_i, b_i \in \mathbb{F}_{13}$. Messages $W_1, W_2$ are defined in terms of these symbols as follows.
{%\setlength{\mathindent}{0cm}
\begin{eqnarray}
W_1 = (a_1; a_2; a_3; a_4), W_2 = (b_1; b_2; b_3; b_4)
\end{eqnarray}}
The storage is specified as
\begin{eqnarray}
{\mbox{Server 1}}:&&W_{11} = (a_1; a_2), W_{21} = (b_1; b_2) \\
{\mbox{Server 2}}:&&W_{12} = (a_3; a_4), W_{22} = (b_3; b_4) \\
{\mbox{Server 3}}:&&W_{13} = (\alpha_1; \alpha_2), W_{23} = (\beta_1; \beta_2) \\
{\mbox{Server 4}}:&&W_{14} = (\alpha_3; \alpha_4), W_{24} = (\beta_3; \beta_4)
\end{eqnarray}
where $\alpha_1, \alpha_2, \alpha_3, \alpha_4, \beta_1, \beta_2, \beta_3, \beta_4$ are obtained as follows.
\begin{eqnarray}
\alpha_1 = 3a_1 + 2a_2 + 4a_3 + a_4, &&\beta_1 = 3b_1 + 2b_2 + 4b_3 + b_4 \notag\\
\alpha_2 = 2a_1 +3a_2 + a_3 + 4a_4, &&\beta_2 = 2b_1 +3b_2 + b_3 + 4b_4 \notag\\
\alpha_3 = 3a_1 + 12a_2 + 4a_3 + 6a_4, &&\beta_3 = 3b_1 + 12b_2 + 4b_3 + 6b_4 \notag\\
\alpha_4 = 12a_1 + 3a_2 + 6a_3 + 4a_4, &&\beta_4 = 12b_1 + 3b_2 + 6b_3 + 4b_4
\end{eqnarray}
Note that each server stores 2 symbols of each message and the storage at any 2 servers is just enough to recover both messages (MDS storage property is satisfied).

We now describe the PIR scheme. $\mathcal{F}$ is a uniform random variable in $[1:64]$. The user's query is uniform over $64$ choices and is specified by Table 2. Note that the queries to servers $1$, $2$ and $3$ are the same, regardless of the value of $\theta$ and the query to Server $4$ is a deterministic function of that to servers $1$, $2$ and $3$.

\begin{table}[h]
\centering
\caption{\small The Scheme for MDS-TPIR with $(K,N,T,K_c) = (2,4,3,2)$.}
\begin{tabular}{|c|c|c|c|c|c|c|}\hline
\mbox{Prob.}&$Q_1^{[\theta]}$~(\mbox{Server $1$})&$Q_2^{[\theta]}$~(\mbox{Server $2$})&$Q_3^{[\theta]}$~(\mbox{Server $3$})&$Q_4^{[1]}$~(\mbox{Server 4})&$Q_4^{[2]}$~(\mbox{Server 4})\\\hline
$1/64$ &``$a_{i_1},b_{j_1}$''&``$a_{i_2},b_{j_2}$''&``$\alpha_{i_3}, \beta_{j_3}$''&``$\alpha_{i_4} + \beta_{j_4}$ & ``$\alpha_{i_4'} + \beta_{j_4'}$''\\
\hline
\end{tabular}\\
{$i_1, j_1, i_3, j_3$ are i.i.d. and uniform in $\{1,2\}$. $i_2, j_2$ are i.i.d. and uniform in $\{3,4\}$. \\
$i_4, j_4, i_4', j_4'$ are determined as follows.}
\begin{eqnarray*}
(i_1, i_2, i_3) = (1,3,1) \Rightarrow i_4 =  4, i_4' = 3, && (j_1, j_2, j_3) = (1,3,1) \Rightarrow j_4 =  3, j_4' = 4\\
(i_1, i_2, i_3) = (1,3,2) \Rightarrow i_4 =  3, i_4' = 4, && (j_1, j_2, j_3) = (1,3,2) \Rightarrow j_4 =  4, j_4' = 3\\
(i_1, i_2, i_3) = (1,4,1) \Rightarrow i_4 =  3, i_4' = 4, && (j_1, j_2, j_3) = (1,4,1) \Rightarrow j_4 =  4, j_4' = 3\\
(i_1, i_2, i_3) = (1,4,2) \Rightarrow i_4 =  4, i_4' = 3, && (j_1, j_2, j_3) = (1,4,2) \Rightarrow j_4 =  3, j_4' = 4\\
(i_1, i_2, i_3) = (2,3,1) \Rightarrow i_4 =  3, i_4' = 4, && (j_1, j_2, j_3) = (2,3,1) \Rightarrow j_4 =  4, j_4' = 3\\
(i_1, i_2, i_3) = (2,3,2) \Rightarrow i_4 =  4, i_4' = 3, && (j_1, j_2, j_3) = (2,3,2) \Rightarrow j_4 =  3, j_4' = 4\\
(i_1, i_2, i_3) = (2,4,1) \Rightarrow i_4 =  4, i_4' = 3, && (j_1, j_2, j_3) = (2,4,1) \Rightarrow j_4 =  3, j_4' = 4\\
(i_1, i_2, i_3) = (2,4,2) \Rightarrow i_4 =  3, i_4' = 4, && (j_1, j_2, j_3) = (2,4,2) \Rightarrow j_4 =  4, j_4' = 3
\end{eqnarray*}
\end{table}

The key to the scheme is that the undesired symbol downloaded from Server $4$ is known from that downloaded from servers $1$, $2$ and $3$, while desired symbols are all independent. This observation is formalized in the following lemma.
\begin{lemma}\label{lemma:align2}
For all values of $i_1, i_2, i_3, i_4, i_4', j_1, j_2, j_3, j_4, j_4'$ in Table 2, we have
\begin{eqnarray}
\dim(a_{i_1}, a_{i_2}, \alpha_{i_3}, \alpha_{i_4}) = 4,&& \dim(a_{i_1}, a_{i_2}, \alpha_{i_3}, \alpha_{i_4'}) = 3\label{eq:align3} \\
\dim(b_{j_1}, b_{j_2}, \beta_{j_3}, \beta_{j_4}) = 3,&& \dim(b_{j_1}, b_{j_2}, \beta_{j_3}, \beta_{j_4'}) = 4 \label{eq:align4}
\end{eqnarray}
\end{lemma}
Lemma \ref{lemma:align2} is proved by brute force, i.e., verifying (\ref{eq:align3}) and (\ref{eq:align4}) hold for each case.

We show that the scheme is both correct and private.
The schemes is correct because as Lemma \ref{lemma:align2} has proved, the $4$ undesired symbols only have dimension $3$ and it is easy to see that the $3$ undesired symbols in answers from the first $3$ servers have dimension 3. Therefore, from the answers $A_1^{[k]}, A_2^{[k]}, A_3^{[k]}$, we always know the undesired symbol in $A_4^{[k]}$. Subtracting the undesired symbol out from $A_4^{[k]}$, we obtain the desired symbol interference freely. Lemma \ref{lemma:align2} has proved that the $4$ desired symbols are independent such that we can recover the desired message.
The scheme is private because it is easy to verify that for any $3$ servers, the queries are identically distributed no matter which message is desired and then the privacy condition (\ref{privacy}) is satisfied.

The scheme downloads $2$ symbols from Server $1$, Server $2$ and Server $3$ each, and $1$ symbol from Server $4$. It retrieves $4$ desired message symbols. Therefore the rate is $4/7$.

Let us conclude this example with the observation that this MDS-TPIR instance with $(K,N,T,K_c)$ $=$ $(2,4,3,2)$ is not covered by Theorem \ref{thm:class},  but we were still able to find its capacity. Let us also note that we are able to cast this example into a similar framework as Theorem \ref{thm:class} and prove the existence of  PIR schemes that achieve the same capacity for the $(x,y)\rightarrow(x,y,x+y,x+2y)$ MDS storage code, subject to the assumption of a sufficiently large finite field. The details are repetitive, and therefore omitted. However, we believe this example may provide useful insights for further generalizations.}

 \subsection{Achievability Proof for Theorem \ref{thm:class} when $T = 2$}\label{sec:ach_proof2}
The proof for the general setting (arbitrary $N$) follows the same route as the $N = 4$ example presented earlier.
We assume that each message is comprised of $L=N(N-1)$ independent symbols from a sufficiently large finite field $\mathbb{F}_p$. 

\subsubsection{Storage Code}
The $(N-1,N)$ MDS storage code is as follows. 
\begin{eqnarray}
W_{kn} &\in& \mathbb{F}_{p}^{N \times 1}, k \in [1:2], n\in[1:N]\\
W_k &=& (W_{k1}; W_{k2}; \cdots; W_{k(N-1)}) \in \mathbb{F}_p^{L \times 1}\\
W_{kN} &=& W_{k1} + W_{k2} + \cdots + W_{k(N-1)}
\end{eqnarray}

\subsubsection{Construction of Queries}
The query to each server consists of $2(N-1)$ vectors, the first $N-1$ vectors for $W_1$ ($Q_n^{[k]}(W_1)$) and the last $N-1$ vectors for $W_2$ ($Q_n^{[k]}(W_2)$). The queries and downloads for $W_k, k \in [1:2]$ are described next.

Denote the set of all full rank $N \times N$ matrices over $\mathbb{F}_p$ as $\mathcal{S}_N$. The user privately chooses two matrices $S, S'$, independently and uniformly from $\mathcal{S}_N$. Label the rows of $S$ as $V_1, \cdots, V_N$, and the rows of $S'$ as $\overline{U}_1, \cdots, \overline{U}_{N-2}, U_1, U_2$. Define $\forall n \in [1:N]$
\begin{eqnarray}
\mathcal{V}_n &=& \{V_1, \cdots, V_{n-1}, V_{n+1}, \cdots, V_N\}\\
\mathcal{U}_n &=& \{\overline{U}_1, \cdots, \overline{U}_{N-2}, \widetilde{U}_n\}
\end{eqnarray}
where $\widetilde{U}_n, n \in [1:N]$ are the rows of $\widetilde{U}$, obtained as follows.
\begin{eqnarray}
\widetilde{U} = \mbox{MDS}_{N \times 2} (U_1; U_2) \label{uu2}
\end{eqnarray}
where $\mbox{MDS}_{N \times 2}$ is an $N\times 2$ matrix such that any two of its rows are linearly independent. 

{When $W_k$ is desired}, we have $\forall n,$
{%\setlength{\mathindent}{0.1cm}
\begin{align}
{\mbox{Server}~n}:
Q_n^{[k]}(W_k) &= \mathbb{\pi}_n (\mathcal{V}_n),& A_n^{[k]}(W_k) &= Q_n^{[k]}(W_k) W_{kn}.
\end{align}
}

\indent {\it Desired Symbols Are Independent:} From $A_{1:N}^{[{k}]}(W_k)$, we can recover all $N(N-1)$ symbols of $W_k$. This is easily seen because the storage is an $(N-1, N)$ MDS code, no query dimension is repeated more than $N-1$ times and the matrix $S$ has full rank.

{When $W_k$ is undesired}, we have $\forall n,$
{%\setlength{\mathindent}{0cm}
\begin{align}
{\mbox{Server}~n}: Q_n^{[k^c]}(W_k) &= \mathbb{\pi}_n' (\mathcal{U}_n),& A_n^{[k^c]}(W_k)& = Q_n^{[k^c]}(W_k) W_{kn}.
\end{align}
\indent{\it Interfering Symbols Are Dependent and Have Dimension at most $N(N-1) - (N-2)$:}
Consider the interfering symbols along the common vectors $\overline{U}_i, i \in [1:N-2]$. Note that 
\begin{eqnarray}
%\hspace{1cm}
\overline{U}_i W_{k1} + \cdots + \overline{U}_i W_{k(N-1)} = \overline{U}_i W_{kN} \label{eq:dep_even2}
\end{eqnarray}
Therefore $(N-2)$ interfering symbols are linear combinations of the other $N^2 - 2N + 2$ symbols.

\subsubsection{Combining Answers for Efficient Download}\label{sec:sz}
Based on the queries, each server has $2(N-1)$ symbols, $N-1$ in $W_1$, $A_n^{[k]}(W_1)$ and $N-1$ in $W_2$, $A_n^{[k]}(W_2)$ for a total of $L = N(N-1)$ desired symbols and $L = N(N-1)$ undesired symbols. Note that there are at most $N^2 - 2N + 2 \triangleq I$ independent undesired symbols. Exploiting this fact, we will combine the $2(N-1)$ queried symbols from each server into $(I+L)/N$  symbols to be downloaded by the user. Intuitively, $(L+I)/N$ symbols from each server will give the user a total of $L+I$ symbols, from which he can resolve the $L$ desired and $I$ undesired symbols.

Define the following function that maps $2L/N \in \mathbb{Z}_+$ input symbols to $(L+I)/N \in \mathbb{Z}_+$ output symbols.
\begin{eqnarray}
&&{\color{black}\mathcal{L}^*}(X_{1}, X_2, \cdots, X_{L/N}, Y_1, Y_2, \cdots, Y_{L/N}) \notag\\
&=& (X_1, \cdots, X_{I/N}, Y_1, \cdots, Y_{I/N}, X_{I/N+1} + Y_{I/N+1}, \cdots, X_{L/N} + Y_{L/N}) \label{eq:ll}
\end{eqnarray}

We formalize the combining process in the following lemma.
\begin{lemma}\label{lemma:comb2}
Suppose each server has $L/N$ desired symbols and $L/N$ undesired symbols. Across all servers, the $L$ desired symbols are independent, while the $L$ undesired symbols have dimension at most $I$, i.e., all $L$ undesired symbols can be expressed as linear combinations of symbols in ${\bf s}$, where ${\bf s}$ is a set of $I$ symbols. Further, each server contains $I/N$ distinct symbols in ${\bf s}$.

The desired and undesired symbols are combined to produce the answers as follows.
\begin{eqnarray}
A_n^{[k]} = \mathcal{L}^*(C_n A_n^{[k]}(W_1), C_n A_n^{[k]}(W_{2})) 
\end{eqnarray}
where $C_n$ are {\color{black} deterministic} $L/N \times L/N$ matrices, that are required to satisfy the following two properties. Denote the first $I/N$ rows of $C_n$ as $\overline{C}_n$.
\begin{enumerate}
\item[{\it P1.}] All $C_n$ have full rank.
\item[{\it P2.}] For all $(N-1)!^N$ distinct realizations of $\pi_n', n \in [1:N]$, the $I$  symbols of the undesired message that are directly downloaded ($I/N$ from each server), $\overline{C}_{1} A_1^{[k]}(W_{k^c})$, $\overline{C}_{2} A_2^{[k]}(W_{k^c})$, $\cdots$, $\overline{C}_{N} A_N^{[k]}(W_{k^c})$ are independent in variables in ${\bf s}$. 
\end{enumerate}

Then we have the following claim.

\noindent {\it Claim.} The $C_n$ satisfying the two required properties exist over $\mathbb{F}_p$ for a sufficiently large prime $p$.\footnote{In fact, the properties are generic, i.e., they are satisfied by almost all matrices over large fields.}
\end{lemma}

\proof This proof of existence will use Schwartz-Zippel lemma \cite{Schwartz, Zippel} about the roots of a polynomial. The variables for the polynomial are the coefficients of the $C_n$ matrices. {\color{black}Let us start with an arbitrary choice of $\pi_n', n \in [1:N]$}.
Since all $A_n^{[k]}(W_{k^c})$ can be expressed in terms of the $I$ symbols in %\footnote{$H({\bf s}|U_{0:9})=8$ in $p$-ary units.} of 
the vector ${\bf s} %= (U_0 {\bf x}; U_6 {\bf x}$; $U_0 {\bf y}$; $U_9 {\bf y}$; $U_1  ({\bf x}+{\bf y})$; $U_3  ({\bf x}+{\bf y})$; $U_2  ({\bf x}+2{\bf y})$; $U_4  ({\bf x}+2{\bf y}))
$ with constant coefficients, %(refer to (\ref{eq:dependent})), 
we can express
\begin{eqnarray}
(\overline{C}_{1} A_1^{[k]}(W_{k^c}); \cdots; \overline{C}_{N} A_N^{[k]}(W_{k^c})) = \mathcal{C}_{I\times I} {\bf s}
\end{eqnarray}
Now consider the polynomial given by the determinant of $\mathcal{C}$. This is not the zero polynomial\footnote{A polynomial is a zero polynomial if all its coefficients are zero.} because we can easily assign values to $\overline{C}_n$  to make $\mathcal{C} = I$, the identity matrix. This is  because the queried symbols from each server include $I/N$ distinct symbols in ${\bf s}$. 

{\color{black}Next do the same for \emph{every} realization of $\pi_n', n \in [1:N]$. As there are $N$ permutations involved, and each can take $(N-1)!$ different values, so we have a total of $(N-1)!^N$ different possibilities. We will consider each of them separately. Each time we find a different $\mathcal{C}$, which gives us a different non-zero polynomial.}

Next consider the determinant of each $C_n$. This gives us another $N$ non-zero polynomials.

For each of these $(N-1)!^N + N$ polynomials, Schwartz-Zippel lemma guarantees that a uniformly random choice of $C_n$ produces a non-zero evaluation with high probability over a large field (probability approaching $1$ as $p \rightarrow \infty$). Since the intersection of finite number of high probability events is also a high probability event, there must exist a realization of $C_n$ over a large field for which all $(N-1)!^N + N$ polynomials simultaneously evaluate to non-zero values, i.e., a realization that satisfies both properties. Hence, the claim is true.

%Finally, consider the polynomial obtained by multiplying all $5$ polynomials. A product of non-zero polynomials is also a non-zero polynomial. Let the field size be larger than the degree of this polynomial. Then the Schwartz-Zippel lemma guarantees a non-zero probability that the polynomial evaluates to a non-zero value for a uniformly random choice of $C_n$.\footnote{In fact, as the field size increases, randomly chosen $C_n$ produce a non-zero evaluation of the polynomial with a probability that approaches $1$.} Since the probability is not zero,  there exists a realization of $C_n$ matrices for which the polynomial is non-zero, i.e., a realization that satisfies both properties. Hence, we have proved the claim.
\hfill\QED
}

Next we prove that the  scheme retrieves the desired message,  and that it is $T$ private.

%\subsection
\subsubsection{The Scheme is Correct (Retrieves Desired Message)}
Note that from (\ref{eq:dep_even2}), independent undesired message symbols distribute evenly across the databases, such that Lemma \ref{lemma:comb2} applies. 
Note that the first $2I/N$ variables in the output of the $\mathcal{L}^*$ function are obtained directly, i.e., $\overline{C}_{1} A_1^{[k]}(W_1)$, $\overline{C}_{2} A_2^{[k]}(W_1)$, $\cdots$, $\overline{C}_{N} A_N^{[k]}(W_1)$ and $\overline{C}_{1} A_1^{[k]}(W_2)$, $\overline{C}_{2} A_2^{[k]}(W_2)$, $\cdots$, $\overline{C}_{N} A_N^{[k]}(W_2)$ are all directly recovered. By property {\it P2} of $C_n$, $\overline{C}_{1} A_1^{[k]}(W_{k^c})$, $\overline{C}_{2} A_2^{[k]}(W_{k^c})$, $\cdots$, $\overline{C}_{N} A_N^{[k]}(W_{k^c})$ are linearly independent. Since we have recovered $I$ independent dimensions of interference, and interference only spans at most $I$ dimensions, all interference is recovered and eliminated. Further, since the $L$ desired symbols are independent and since the $C_n$ matrices have full rank, the user is able to recover the $L$ desired message symbols after the interference symbols are recovered and subtracted from the downloaded equations. Therefore the scheme is correct {\color{black} with zero error}.

%\subsection
\subsubsection{The Scheme is Private (to any $T=2$ Colluding Servers)}
To prove that the scheme is $T = 2$ private (refer to (\ref{privacy})), it suffices to show that the queries for any $2$ servers are identically distributed, regardless of which message is desired. Since each query is made up of $2(N-1)$ vectors, $N-1$ for each message and the vectors for $W_1$ and the vectors for $W_2$ are generated independently, it suffices to prove that the vectors for one message (say $W_k$) are identically distributed, i.e.,
{%\setlength{\mathindent}{0cm}
\begin{eqnarray}
\left(Q_{n_1}^{[k]}(W_k), Q_{n_2}^{[k]}(W_k)\right) \sim \left(Q_{n_1}^{[k^c]}(W_k), Q_{n_2}^{[k^c]} (W_k)\right), 
~\forall n_1, n_2  \in [1:4], n_1 < n_2 \label{space_privacy22}
\end{eqnarray}}
Note that  
{\color{black} \begin{eqnarray}
\left(Q_{n_1}^{[k]}(W_k), Q_{n_2}^{[k]}(W_k)\right) = \left(\mathbb{\pi}_{n_1}(\mathcal{V}_{n_1}), \mathbb{\pi}_{n_2}(\mathcal{V}_{n_2}) \right) \\
\left(Q_{n_1}^{[k^c]}(W_k), Q_{n_2}^{[k^c]}(W_k)\right) = \left(\mathbb{\pi}_{n_1}' (\mathcal{U}_{n_1}), \mathbb{\pi}_{n_2}' (\mathcal{U}_{n_2}) \right)
\end{eqnarray}}
Therefore, to prove (\ref{space_privacy22}) it suffices to show the following.
\begin{eqnarray}
\big( {V_{1}}, \cdots, {V_{i_{n_1-1}}, V_{i_{n_1+1}}}, \cdots, {V_{i_{n_2-1}}, V_{i_{n_2 +1}}}, \cdots, V_N, V_{i_{n_2}}, V_{i_{n_1}} \big) \sim \big( \overline{U}_1, \cdots, \overline{U}_{N-2}, \widetilde{U}_{n_1}, \widetilde{U}_{n_2} \big)
\label{p522}
\end{eqnarray}
Because $S$ is uniformly chosen from the set of all full rank matrices, we have
\begin{eqnarray}
\big( {V_{1}}, \cdots, {V_{i_{n_1-1}}, V_{i_{n_1+1}}}, \cdots, {V_{i_{n_2-1}}, V_{i_{n_2 +1}}}, \cdots, V_N, V_{i_{n_2}}, V_{i_{n_1}} \big)  \sim  S
\label{p222}
\end{eqnarray}
Recall that $S = (V_1, \cdots, V_N)$.
Next we note that there is a bijection between 
\begin{eqnarray}
 \big( \overline{U}_1, \cdots, \overline{U}_{N-2}, \widetilde{U}_{n_1}, \widetilde{U}_{n_2} \big) &\leftrightarrow& S'
\end{eqnarray}
because of (\ref{uu2}) so that there is a bijection between $\widetilde{U}_{n_1}, \widetilde{U}_{n_2}$ and $U_1, U_2$. Recall that $S' =  \big( \overline{U}_1, \cdots, \overline{U}_{N-2}, {U}_1, {U}_2 \big)$.
Now as $S'$ is uniform over all full rank matrices, $\big( \overline{U}_1, \cdots, \overline{U}_{N-2}, \widetilde{U}_{n_1}, \widetilde{U}_{n_2} \big)$ is also uniform over all full rank matrices,
\begin{eqnarray}
 \big( \overline{U}_1, \cdots, \overline{U}_{N-2}, \widetilde{U}_{n_1}, \widetilde{U}_{n_2} \big) &\sim& S'
 \label{p322}
\end{eqnarray}
Finally, we note that $S$ and $S'$ are identically distributed, so we have
\begin{eqnarray}
S \sim S'
\label{p422}
\end{eqnarray}
Combining (\ref{p222}), (\ref{p322}) and (\ref{p422}), we arrive at (\ref{p522}) and (\ref{space_privacy22}).

\subsubsection{Rate Achieved is $(N^2 - N)/(2N^2 - 3N + 2)$}
The rate achieved is $(N^2 - N)/(2N^2 - 3N + 2)$, because we download $2N^2 - 3N + 2$ symbols in total and the desired message size is $N(N-1)$ symbols.

{
\subsection{Achievability Proof of Theorem \ref{thm:class} when $T >2$}\label{sec:ach_proof}
The proof for the general setting follows the same route as the $N = 4, T = 3$ example presented earlier.
We assume that each message is comprised of $L=N(N-1)$ independent symbols from a sufficiently large finite field $\mathbb{F}_p$. 

\subsubsection{Storage Code}
The $(N-1,N)$ MDS storage code is as follows. 
\begin{eqnarray}
W_{kn} &\in& \mathbb{F}_{p}^{N \times 1}, k \in [1:2], n\in[1:N]\\
W_k &=& (W_{k1}; W_{k2}; \cdots; W_{k(N-1)}) \in \mathbb{F}_p^{L \times 1}\\
W_{kN} &=& W_{k1} + W_{k2} + \cdots + W_{k(N-1)}
\end{eqnarray}

\subsubsection{Construction of Queries}
The query to each server consists of two vector spaces, one for $W_1$ (span of the rows of $Q_n^{[k]}(W_1)$) and one for $W_2$ (span of the rows of  $Q_n^{[k]}(W_2)$). The queries and downloads for $W_k, k \in [1:2]$ are described next.

Denote the set of all full rank $N \times N$ matrices over $\mathbb{F}_p$ as $\mathcal{S}_N$. The user privately chooses two matrices $S, S'$, independently and uniformly from $\mathcal{S}_N$. Label the rows of $S$ as $V_1, \cdots, V_N$, and the rows of $S'$ as $\overline{U}_1, \cdots, \overline{U}_{N-T}, U_1, \cdots, U_T$. Define $\forall n \in [1:N]$
\begin{eqnarray}
\mathcal{V}_n &=& \{V_1, \cdots, V_{n-1}, V_{n+1}, \cdots, V_N\}\\
\mathcal{U}_n &=& \{\overline{U}_1, \cdots, \overline{U}_{N-T}, \widetilde{U}_{(n-1)(T-1)+1}, \cdots, \widetilde{U}_{n(T-1)}\}
\end{eqnarray}
where $\widetilde{U}_1, \cdots, \widetilde{U}_{N(T-1)}$ are the rows of $\widetilde{U}$, obtained as follows.
\begin{eqnarray}
\widetilde{U} = {P} (U_1; \cdots; U_T) \label{uu}
\end{eqnarray}
${P}$ is a deterministic $N(T-1) \times T$ matrix that is chosen in such a way that it allows a bijective mapping between the $\mathcal{U}_n$ or $\mathcal{V}_n$ spaces that may be observed by any set of up to $T$ colluding servers. 
%The specification of the properties of this matrix, as well as its proof of existence is relegated to Section \ref{sec:ach_proof}. 
Intuitively, the only requirement on this matrix is that it is sufficiently `generic', so that almost all $N(T-1) \times T$ matrices over large finite fields are acceptable. Here unlike the previous example where we explicitly construct the matrix $P$, we will specify (later) the properties of this matrix and prove that such a matrix exists.

{When $W_k$ is desired}, we have $\forall n,$
{%\setlength{\mathindent}{0.1cm}
\begin{eqnarray}
{\mbox{Server}~n}:
Q_n^{[k]}(W_k) &= \mathbb{B} (\mathcal{V}_n), A_n^{[k]}(W_k) = Q_n^{[k]}(W_k) W_{kn}.
\end{eqnarray}
}
\indent {\it Desired Symbols Are Independent:} From $A_{1:N}^{[{k}]}(W_k)$, we can recover all $N(N-1)$ symbols of $W_k$. This is easily seen because the storage is an $(N-1, N)$ MDS code and the matrix $S$ has full rank.

{When $W_k$ is undesired}, we have $\forall n,$
{%\setlength{\mathindent}{0cm}
\begin{eqnarray}
{\mbox{Server}~n}: Q_n^{[k^c]}(W_k) &= \mathbb{B} (\mathcal{U}_n), A_n^{[k^c]}(W_k) = Q_n^{[k^c]}(W_k) W_{kn}.
\end{eqnarray}
\indent{\it Interfering Symbols Are Dependent and Have Dimension at most $N(N-1) - (N-T)$:}
Consider the interfering symbols along the common vectors $\overline{U}_i, i \in [1:N-T]$. Note that 
\begin{eqnarray}
%\hspace{1cm}
\overline{U}_i W_{k1} + \cdots + \overline{U}_i W_{k(N-1)} = \overline{U}_i W_{kN} \label{eq:dep_even}
\end{eqnarray}
Therefore $(N-T)$ interfering symbols are linear combinations of the other $N^2 - 2N + T$ symbols. 

\subsubsection{Combining Answers for Efficient Download}
{\color{black}The idea of combining is the same as the $T = 2$ setting. That is, we will combine the $2(N-1)$ queried symbols from each server into $(2N^2-3N+T)/N = (L+I)/N$ symbols to be downloaded by the user. We will use the same combining function $\mathcal{L}^*$ defined in (\ref{eq:ll}). The difference lies in the combining matrices $C_n$. For $T = 2$, $C_n$ are deterministic and the scheme has zero-error, while here $C_n$ are random and the scheme has $\epsilon$-error, with $\epsilon$ approaching zero as the message size approaches infinity. The combining process is described in the following lemma, which corresponds to Lemma \ref{lemma:comb2} (with differences brought by random $C_n$ accounted).}

%Based on the queries, each server has $2(N-1)$ symbols, $N-1$ in $W_1$, $A_n^{[k]}(W_1)$ and $N-1$ in $W_2$, $A_n^{[k]}(W_2)$ for a total of $N(N-1)$ desired symbols and $N(N-1)$ undesired symbols. Note that there are at most $N^2 - 2N + T \triangleq I$ independent undesired symbols. Exploiting this fact, we will combine the $2(N-1)$ queried symbols from each server into $(2N^2-3N+T)/N$  symbols to be downloaded by the user. Intuitively, $(2N^2-3N+T)/N$ symbols from each server will give the user a total of $2N^2-3N+T$ symbols, from which he can resolve the $N(N-1)$ desired and $I$ undesired symbols.

%Define the following function that maps $2L/N \in \mathbb{Z}_+$ input symbols to $(L+I)/N \in \mathbb{Z}_+$ output symbols.
%{\begin{eqnarray}
%&&{\color{black}\mathcal{L}^*}(X_{1}, X_2, \cdots, X_{L/N}, Y_1, Y_2, \cdots, Y_{L/N}) \notag\\
%&=& (X_1, \cdots, X_{I/N}, Y_1, \cdots, Y_{I/N}, X_{I/N+1} + Y_{I/N+1}, \cdots, X_{L/N} + Y_{L/N})
%\end{eqnarray}
%}

%In order to have a general statement, we formalize the combining process in the following lemma.
\begin{lemma}\label{lemma:comb}
Suppose each server has $L/N$ desired symbols and $L/N$ undesired symbols from $\mathbb{F}_p$. Across all servers, the $L$ desired symbols are independent, while the $L$ undesired symbols have dimension at most $I$, i.e., all $L$ undesired symbols can be expressed as linear combinations of symbols in ${\bf s}$, where ${\bf s}$ is a set of $I$ symbols. Further, each server contains $I/N$ distinct symbols in ${\bf s}$.

The desired and undesired symbols are combined to produce the answers as follows.
\begin{eqnarray}
A_n^{[k]} = \mathcal{L}^*(C_n A_n^{[k]}(W_1), C_n A_n^{[k]}(W_{2})) 
\end{eqnarray}
where $C_n$ are {\color{black} random} $L/N \times L/N$ matrices, that are required to satisfy the following two properties. Denote the first $I/N$ rows of $C_n$ as $\overline{C}_n$.
\begin{enumerate}
\item[{\it P1.}] All $C_n$ are full rank.
\item[{\it P2.}] The $I$  symbols of the undesired message that are directly downloaded ($I/N$ from each server), $\overline{C}_{1} A_1^{[k]}(W_{k^c})$, $\overline{C}_{2} A_2^{[k]}(W_{k^c})$, $\cdots$, $\overline{C}_{N} A_N^{[k]}(W_{k^c})$ are independent in variables in ${\bf s}$. 
\end{enumerate}

Then the following claim must be true.

\vspace{0.04in}
{\color{black}\noindent {\it Claim.} The probability that $C_n, n\in[1:N]$ with each element chosen independently and uniformly over $\mathbb{F}_p$, satisfy the two required properties, approaches $1$ as $p\rightarrow\infty$.} %The $C_n$ satisfying the two required properties exist over a sufficiently large field. 
\vspace{0.04in}
\end{lemma}

{\it Proof:}
Without loss of generality, we assume that $I/N$ is an integer. There is no loss of generality because if $I/N$ is not an integer, we may repeat the scheme a number of times (say $M$) such that $IM/N$ becomes an integer.

The proof relies on Schwartz-Zippel lemma \cite{Schwartz, Zippel} about the roots of a polynomial. The variables for the polynomial are the coefficients of the $C_n$ matrices. 
Consider an arbitrary realization of the query spaces $\mathcal{U}_n$. Generate uniformly random $C_n$, independent of $\mathcal{U}_n$.
%Consider the $I$ symbols $\overline{C}_{1} A_1^{[k]}(W_{k^c})$, $\overline{C}_{2} A_2^{[k]}(W_{k^c})$, $\cdots$, $\overline{C}_{N} A_N^{[k]}(W_{k^c})$. As $C_n$ are chosen uniformly over $\mathbb{F}_p$ and $C_n$ are independent of ${\bf s}$, the $I$ symbols have full rank almost surely. %This is because the determinant polynomial is not the zero polynomial as we can easily assign values to $\overline{C}_n$  to make $\mathcal{C} = I$, the identity matrix. This is because each server contains $I/N$ distinct symbols in ${\bf s}$. So by the Schwartz-Zippel lemma, a non-zero polynomial evaluates to a non-zero value with probability approaching $1$ as the field size $p$ increases and $C_n$ are chosen uniformly over $\mathbb{F}_p$.
%(Have not fully understood the proof. For a given $\mathcal{U}_n$, with large $p$, random $C_n$ will work almost surely. But still we have a lot of $\mathcal{U}_n$, can we guarantee that for all $\mathcal{U}_n$, the probability of error still approaches 0? Seems we have the same issue as the zero error proof?)
Given $\mathcal{U}_n, n\in[1:N]$, since all $A_n^{[k]}(W_{k^c})$ can be expressed in terms of the $I$ symbols of the vector ${\bf s}$ with constant coefficients, we can express
\begin{eqnarray}
(\overline{C}_{1} A_1^{[k]}(W_{k^c}); \cdots; \overline{C}_{N} A_N^{[k]}(W_{k^c})) = \mathcal{C}_{I \times I} {\bf s}
\end{eqnarray}
Now consider the polynomial given by the determinant of $\mathcal{C}$. This is not the zero polynomial because we can easily assign values to $\overline{C}_n$  to make $\mathcal{C} = I$, the identity matrix. This is  because each server contains $I/N$ distinct symbols in ${\bf s}$. By the Schwartz-Zippel lemma, a non-zero polynomial evaluates to a non-zero value with probability approaching $1$ as the field size $p$ increases and $C_n$ are chosen uniformly over $\mathbb{F}_p$. Therefore Property $P2$ is satisfied with high probability.

Next consider the determinant of each $C_n$. This gives us another $N$ non-zero polynomials. {\color{black} When we choose $C_n$ uniformly, the determinant of $C_n$ is not zero almost surely for large $p$, so that $C_n$ have full rank and Property $P1$ is satisfied with high probability.

Now, because Property $P1$ and $P2$ are each satisfied with probability approaching $1$, the probability that the two are simultaneously satisfied also approaches $1$ (union bound). 
Since this is true conditioned on every possible realization of $\mathcal{U}_n, n\in[1:N]$,  it is also true  unconditionally.

%{\it Remark: The difference from Theorem \ref{thm:disprove} is that $C_n$ are now different for different realizations of $\mathcal{U}_n$. Note that the number of possible realizations of $\mathcal{U}_n$ increases with the field size $p$ (in particular, is at least comparable with $p$). Therefore, if we follow the approach of that with $T=2$, we can no longer guarantee the existence of common $C_n$ that work for all cases. This is the reason that here we generate independent $C_n$ for different realizations of $\mathcal{U}_n$. As a price to pay, we have a $\epsilon$-error scheme, instead of a zero-error one.}
}
%By the Schwartz-Zippel lemma, each non-zero polynomial evaluates to a non-zero value with probability approaching $1$ as the field size $p$ increases and $C_n$ are chosen uniformly over $\mathbb{F}_p$. By the union bound, the probability that all of them simultaneously evaluate to a non-zero value also approaches $1$. In particular, the probability is not zero. Since the probability is not zero,  there exists a realization of $C_n$ matrices that satisfies both properties. 
\hfill\QED
%Finally, consider the polynomial obtained by multiplying all $N+1$ polynomials. A product of non-zero polynomials is also a non-zero polynomial. Let the field size be larger than the degree of this polynomial. Then the Schwartz-Zippel lemma guarantees a non-zero probability that the polynomial evaluates to a non-zero value for a uniformly random choice of $C_n$.
%\footnote{In fact, as the field size increases, randomly chosen $C_n$ produce a non-zero evaluation of the polynomial with a probability that approaches $1$.} 

Next we prove that the  scheme retrieves the desired message,  and that it is $T$ private.

%\subsection
\subsubsection{The Scheme is Correct (Retrieves Desired Message)}
Note that from (\ref{eq:dep_even}), independent undesired message symbols distribute evenly across the databases, such that Lemma \ref{lemma:comb} applies. 
Note that the first $2I/N$ variables in the output of the $\mathcal{L}^*$ function are obtained directly, i.e., $\overline{C}_{1} A_1^{[k]}(W_1)$, $\overline{C}_{2} A_2^{[k]}(W_1)$, $\cdots$, $\overline{C}_{N} A_N^{[k]}(W_1)$ and $\overline{C}_{1} A_1^{[k]}(W_2)$, $\overline{C}_{2} A_2^{[k]}(W_2)$, $\cdots$, $\overline{C}_{N} A_N^{[k]}(W_2)$ are all directly recovered. By property {\it P2} of $C_n$, $\overline{C}_{1} A_1^{[k]}(W_{k^c})$, $\overline{C}_{2} A_2^{[k]}(W_{k^c})$, $\cdots$, $\overline{C}_{N} A_N^{[k]}(W_{k^c})$ are linearly independent {\color{black} with probability approaching $1$ as $p\rightarrow\infty$}. Since we have recovered $I$ independent dimensions of interference, and interference only spans at most $I$ dimensions, all interference is recovered and eliminated. Further, since the $L$ desired symbols are independent and since the $C_n$ matrices have full rank, the user is able to recover the $L$ desired message symbols after the interference symbols are recovered and subtracted from the downloaded equations. Therefore the scheme is correct {\color{black} with a probability of error} $\epsilon$ that approaches $0$ as the field size $p$ approaches infinity. Note that since each message is comprised of $L$ independent and uniformly random symbols in $\mathbb{F}_p$, as $p$ approaches infinity, the size of each message also approaches infinity. So, given any $\epsilon>0$, we can find a sufficiently large $p$, and a correspondingly large message size value such that the probability of error of the scheme described above, is less than $\epsilon$.

%\subsection
\subsubsection{The Scheme is Private (to any $T$ Colluding Servers)}

To prove that the scheme is $T$ private (refer to (\ref{privacy})), it suffices to show that the queries for any $T$ servers are identically distributed, regardless of which message is desired. Since each query is made up of two vector spaces, one for each message and the two vector spaces are generated independently, it suffices to prove that the query spaces for one message (say $W_k$) are identically distributed whether it is desired or undesired. Consider an index set $\mathcal{T} = \{i_1, i_2, \cdots, i_T\} \subset [1:N]$ such that $i_1 < i_2 < \cdots < i_T$. For all $\mathcal{T}$, we require 
\begin{eqnarray}
\left(Q_{i_1}^{[k]}(W_k), \cdots, Q_{i_T}^{[k]}(W_k) \right) &\sim& \left(Q_{i_1}^{[k^c]}(W_k), \cdots, Q_{i_T}^{[k^c]}(W_k) \right) \\
\Longleftrightarrow ~~~~ \left(\mathbb{B}(\mathcal{V}_{i_1}), \cdots, \mathbb{B}(\mathcal{V}_{i_T}) \right) &\sim& \left( \mathbb{B}(\mathcal{U}_{i_1}), \cdots, \mathbb{B}(\mathcal{U}_{i_T})  \right) \label{space_privacy_general}
\end{eqnarray}
Note that 
\begin{eqnarray}
&& \left(\mathbb{B}(\mathcal{V}_{i_1}), \mathbb{B}(\mathcal{V}_{i_2}), \cdots, \mathbb{B}(\mathcal{V}_{i_T}) \right) \notag\\
&=& \left(\mathbb{B}(\{V_{\mathcal{T}^c}, V_{i_2}, \cdots, V_{i_T}\}), \mathbb{B}(\{V_{\mathcal{T}^c}, V_{i_1}, V_{i_3}, \cdots, V_{i_T}\}), \mathbb{B}(\{V_{\mathcal{T}^c}, V_{i_1}, \cdots, V_{i_{T-1}}\}) \right) \label{eq:vijl_general}
\end{eqnarray}
Next we transform the spaces on the RHS of (\ref{space_privacy_general}) to the form that is the same as (\ref{eq:vijl_general}). 
To do this, we require the matrix $P$ to satisfy the following properties.
\begin{enumerate}
\item[{\it P1.}] For all $\mathcal{T}^* =\{j_1,j_2,\cdots,j_{T-1}\}\subset [1:N], |\mathcal{T}^*| = T-1$, $j_1<j_2<\cdots <j_{T-1}$, there exists a function $m_{\mathcal{T}^*}(P)$ that returns a non-zero vector which lies simultaneously in the spans of each of {\color{black}$P_{j_t}\triangleq P((j_t-1)(T-1)+1:j_t(T-1),:)$}, $t\in[1:T-1]$. %{\color{red}(Please check if this notation is defined.)} 
Note that $m_{\mathcal{T}^*}(P)$ is a $1 \times T$ row vector that only depends on $P$ (it does not depend on $U$).
%\item[{\it P1.}] For all $\mathcal{T}^* =\{j_1,j_2,\cdots,j_{T-1}\}\subset [1:N], |\mathcal{T}^*| = T-1$, we can find one unique vector $U_{\mathcal{T}^*} \triangleq m_{\mathcal{T}^*}(P) U$  in the span of all $\mathbb{B}(\mathcal{U}_i), i \in \mathcal{T}^*$. Note that $m_{\mathcal{T}^*}(P)$ is a $1 \times T$ row vector that only depends on $P$ (it does not depend on $U$).
\item[{\it P2.}] For each $\mathcal{T} = \{i_1, i_2, \cdots, i_T\} \subset [1:N]$, the vectors $m_{\mathcal{T}^*}(P), \forall \mathcal{T}^* \subset \mathcal{T}, |\mathcal{T}^*| = T-1$ (found in {\it P1}) are linearly independent. Equivalently, we require the following $T \times T$ matrix to have full rank.
\begin{eqnarray}
P_{\mathcal{T}} \triangleq (m_{\{i_{[1:T]/\{T\}}\}}(P); m_{\{i_{[1:T]/\{T-1\}}\}}(P); \cdots; m_{\{i_{[1:T]/\{1\}}\}}(P))
\end{eqnarray}
\end{enumerate}

{\it Claim.} The $P$ satisfying the two required properties exists over $\mathbb{F}_p$ for a sufficiently large $p$.

\proof Similar to the proof of existence of $C_n$ matrices presented earlier, this proof of existence will use Schwartz-Zippel lemma \cite{Schwartz, Zippel} about the roots of a polynomial. The variables for the polynomial are the coefficients of the $P$ matrix. Since $P$ is a $N(T-1) \times T$ matrix,  we have a total of $NT(T-1)$ variables. Define a set $\mathcal{P}$ that is comprised of all non-zero polynomials with $NT(T-1)$ variables of $P$ as its variables, and coefficients from $\mathbb{F}_p$.

We first consider Property {\it P1}. Recall that there are  $\binom{N}{T-1}$ choices for $\mathcal{T}^*$. Let us start with an arbitrary choice of $\mathcal{T}^* = \{j_1, j_2, \cdots, j_{T-1}\}$ such that $j_1 < j_2 < \cdots < j_{T-1}$.  The required non-zero vector $m_{\mathcal{T}^*}(P)$ is found as follows.
\begin{eqnarray}
&& m_{\mathcal{T}^*}(P) = H_{1} P_{{j}_1} = H_2 P_{{j}_1} = \cdots = H_{T-1} P_{{j}_{T-1}} \\
&\Rightarrow&
\left[
\begin{array}{cccc}
H_1 & H_2 & \cdots & H_{T-1} 
\end{array}
\right]
\underbrace{\left[
\begin{array}{cccc}
%\underbrace
{P_{{j}_1}}%_{T-1 \times T} 
& P_{{j}_1} & \cdots & P_{{j}_1} \\
-P_{{j}_2} & {\bf 0}  & \cdots & {\bf 0} \\
\vdots & -P_{{j}_3} & \ddots & {\bf 0} \\
{\bf 0} & {\bf 0} & {\bf 0} & -P_{{j}_{T-1}} 
\end{array}
\right]}_{\triangleq P_{\mathcal{J}} %(T-1)^2 \times T(T-2)
} =
\left[
\begin{array}{cccc}
0 & 0 & \cdots & 0 
\end{array}
\right]
\end{eqnarray}
where $P_{{j}_t}, t \in [1:T-1]$ are $(T-1) \times T$ matrices, ${\bf 0}$ is the $(T-1) \times T$ matrix with all elements equal to 0 and $P_{\mathcal{J}}$ is a $(T-1)^2 \times T(T-2)$ matrix.
Note that the left null space of $P_{\mathcal{J}}$ is exactly of one dimension if $P_{\mathcal{J}}$ has full rank. Consider the matrix $P_{\mathcal{J}}^*$, which is a square matrix formed by the last $T(T-2)$ rows of $P_{\mathcal{J}}$. We claim that the determinant of $P_{\mathcal{J}}^*$ is a non-zero polynomial, i.e., $|P_{\mathcal{J}}^*|\in\mathcal{P}$. This is because we can identify a specific choice of $P_{{j}_t}$ such that $|P_{\mathcal{J}}^*|$ is not zero, as follows. We set $P_{{j}_t}$ to be the matrix obtained by inserting an all zero column as the $(T+1- t)^{th}$ column of the $(T-1) \times (T-1)$ identity matrix ${\bf I}_{T-1}$. Equivalently, this means that
\begin{eqnarray}
P_{{j}_t} U &=& (U_1; \cdots; U_{T-t}; U_{T+2-t}; \cdots; U_{T}), t \in [1:T-1]
\end{eqnarray}
Since $U_1, \cdots, U_T$ are independent,  $m_{\mathcal{T}^*}(P) U$ can only be some scaled version of the $U_1$ vector.
This means that $P_{\mathcal{J}}^*$ has full rank (which is also easily verified by plugging the vaules of $P_{{j}_t}$ in $P_{\mathcal{J}}^*$). Therefore, $|P_{\mathcal{J}}^*|\in\mathcal{P}$.
To make $m_{\mathcal{T}^*}(P)$ a function, i.e., to remove ambiguity due to scaling factors,   let us normalize the vector $[H_1, \cdots, H_{T-1}]$ by its first element, $h$, such that this vector is unique (scaling is fixed). Note that {\bf $h\in\mathcal{P}$} because if we use the same special choice of $P_{{j}_t}$ as above, we find that $h = 1$ (non-zero). With normalized $[H_1, \cdots, H_{T-1}]$, we obtain $m_{\mathcal{T}^*}(P)$. Note that each element of $m_{\mathcal{T}^*}(P)$ also belongs to $\mathcal{P}$. 

Now do the same for \emph{every} possible choice of $\mathcal{T}^*$. There are $\binom{N}{T-1}$ possibilities. We will consider each of them separately. Each time we obtain different $|P_{\mathcal{J}}^*|, h \in \mathcal{P}$ and find a different $m_{\mathcal{T}^*}(P)$. Putting all of these together, we have a set of $2 \binom{N}{T-1}$ non-zero polynomials. 

Next consider Property {\it P2}. Similarly, we consider all choices of $\mathcal{T}$ separately. For each choice of $\mathcal{T} = \{i_1, i_2, \cdots, i_T\}$ such that $i_1 < i_2 \cdots < i_T$, we consider the determinant of $P_{\mathcal{T}}$.
This determinant polynomial is non-zero because we may set $P_{{i}_t}, t \in [1:T]$ to be the matrix obtained by inserting an all zero column as the $(T+1- t)^{th}$ column of ${\bf I}_{T-1}$, such that the common vector $m_{\mathcal{T}^*}(P), \forall \mathcal{T}^* \in \mathcal{T}, |\mathcal{T}^*| = T-1$ can be computed explicitly
\begin{eqnarray}
%P_{{i}_1} U &=& (U_1; U_2; \cdots; U_{T-1})\\
%&\cdots&\\
P_{{i}_t} U &=& (U_1; \cdots, U_{T-t}; U_{T+2-t} \cdots; U_{T})\\
%&\cdots&\\
%P_{{i}_T} U &=& (U_2; U_3; \cdots; U_{T})\\
m_{\{i_{[1:T]/\{t\}}\}}(P) &=&  e_{T+1 - t},  \forall t \in [1:T]
%m_{\{i_{[1:T]/\{T-1\}}\}}(P) &=&  \\
%\cdots \\
%m_{\{i_{[1:T]/\{1\}}\}}(P)) = 
\end{eqnarray}
where $e_i$ represents the $1 \times T$ unit row vector with a 1 in the $i^{th}$ location and 0 at all other locations.
Therefore, $P_{\mathcal{T}}$ is an identity matrix and the determinant is 1 (non-zero). With all choices of $\mathcal{T}$, we have another $\binom{N}{T}$ non-zero polynomials.

By Schwartz-Zippel lemma, as the field size grows, for each of the polynomials mentioned above, a uniform choice of $P$ produces a  non-zero evaluation with probability approaching $1$. By the union bound, the probability that all polynomials simultaneously produce a non-zero value also approaches $1$. In particular, for a sufficiently large field this probability is not zero, so there must exist a $P$ matrix that satisfies both properties.\hfill\QED

Because of the two properties, we may equivalently represent $Q_{i_t}^{[k^c]}(W_k), t \in [1:T]$ as
\begin{eqnarray}
Q_{i_t}^{[k^c]}(W_k) =  \mathbb{B}(\mathcal{U}_i) = \mathbb{B}(\{\overline{U}, U_{\{i_{[1:T]/\{1\}}\}}, \cdots, U_{\{i_{[1:T]/\{t-1\}}\}}, U_{\{i_{[1:T]/\{t+1\}}\}}, \cdots, U_{\{ i_{[1:T]/\{T\}}\}} \}),    \label{eq:rep_general}
\end{eqnarray}
%Note that equipped with this representation, $\left( \mathbb{B}(\mathcal{U}_{i_1}), \cdots, \mathbb{B}(\mathcal{U}_{i_T})  \right)$ is now of the same form as $\left(\mathbb{B}(\mathcal{V}_i), \mathbb{B}(\mathcal{V}_j), \mathbb{B}(\mathcal{V}_l) \right)$   and 
We are now ready to prove the privacy condition (\ref{space_privacy_ex}). %Using the representation in (\ref{eq:rep}), we have
\begin{eqnarray}
(\ref{space_privacy_general}) \Longleftrightarrow &&  \left(\mathbb{B}(\{V_{\mathcal{T}^c}, V_{i_2}, \cdots, V_{i_T}\}), \mathbb{B}(\{V_{\mathcal{T}^c}, V_{i_1}, V_{i_3}, \cdots, V_{i_T}\}), \mathbb{B}(\{V_{\mathcal{T}^c}, V_{i_1}, \cdots, V_{i_{T-1}}\}) \right) \notag \\
&\sim& \Big( \mathbb{B}(\{\overline{U}, U_{\{i_{[1:T]/\{2\}}\}}, \cdots, U_{\{i_{[1:T]/\{T\}}\}} \}), \mathbb{B}(\{\overline{U}, U_{\{i_{[1:T]/\{1\}}\}}, U_{\{i_{[1:T]/\{3\}}\}}, \cdots, U_{\{i_{[1:T]/\{T\}}\}} \}), \notag\\
&&~ \cdots, \mathbb{B}(\{\overline{U}, U_{\{i_{[1:T]/\{1\}}\}}, \cdots, U_{\{ i_{[1:T]/\{T-1\}}\}} \})  \Big)
\end{eqnarray}
Therefore, it suffices to show the following.
\begin{eqnarray}
(V_{\mathcal{T}^c}, V_{i_1}, V_{i_2}, \cdots, V_{i_T}) \sim (\overline{U}, U_{\{i_{[1:T]/\{1\}}\}}, U_{\{i_{[1:T]/\{2\}}\}}, \cdots, U_{\{i_{[1:T]/\{T\}}\}}) \label{eq:pp4_general}
\end{eqnarray}
Because $S$ is uniformly chosen from the set of all full rank matrices, we have
\begin{eqnarray}
(V_{\mathcal{T}^c}, V_{i_1}, V_{i_2}, \cdots, V_{i_T}) \sim (V_1, V_2, \cdots, V_N) \label{eq:pp3_general}
\end{eqnarray}
Because of  Property {\it P2}, there is a bijection between 
\begin{eqnarray}
(\overline{U}, U_{\{i_{[1:T]/\{1\}}\}}, U_{\{i_{[1:T]/\{2\}}\}}, \cdots, U_{\{i_{[1:T]/\{T\}}\}}) \leftrightarrow  (\overline{U}, U)
\end{eqnarray}
Now since $S' = (\overline{U}; U)$ is uniform in all full rank matrices, the  bijection implies that  $(\overline{U}$, $U_{\{i_{[1:T]/\{1\}}\}}$, $U_{\{i_{[1:T]/\{2\}}\}}$, $\cdots, U_{\{i_{[1:T]/\{T\}}\}})$ is also uniform in all full rank matrices, i.e.,
\begin{eqnarray}
(\overline{U}, U_{\{i_{[1:T]/\{1\}}\}}, U_{\{i_{[1:T]/\{2\}}\}}, \cdots, U_{\{i_{[1:T]/\{T\}}\}}) \sim  (\overline{U}, U) \label{eq:pp2_general}
\end{eqnarray}
Finally, note that $S$ and $S'$ have the same distribution, so we have
\begin{eqnarray}
(V_1, V_2, \cdots, V_N) \sim (\overline{U}, U) \label{eq:pp1_general}
\end{eqnarray}
Therefore, from (\ref{eq:pp3_general}), (\ref{eq:pp2_general}) and (\ref{eq:pp1_general}), we have proved (\ref{eq:pp4_general}) and (\ref{space_privacy_general}).
\subsubsection{Rate Achieved is $(N^2 - N)/(2N^2 - 3N + T)$}
The rate achieved is $(N^2 - N)/(2N^2 - 3N + T)$, because we download $2N^2 - 3N + T$ symbols in total and the desired message size is $N(N-1)$ symbols. 
}

\subsection{Converse for Arbitrary $K$}\label{sec:converse}
In this section, we consider the information theoretic converse of MDS-TPIR, for two scenarios, 
one with $(K, N, T, K_c) = (K, 4, 2, 2)$ and the other with $(K, N, T, K_c)$ such that $N < T + K_c$. For both scenarios, we provide outer bounds that hold for arbitrary $K$. 

Let us start with two useful lemmas that hold for arbitrary $K, N, T, K_c$.
\begin{lemma}
For all $\mathcal{T} \subset [1:N], |\mathcal{T}| = T$ and $k, k' \in [1:K]$,
\begin{eqnarray}
(A_{\mathcal{T}}^{[k]}, W_1, \cdots, W_K, \mathcal{F}, \mathcal{G}) \sim (A_{\mathcal{T}}^{[k']}, W_1, \cdots, W_K, \mathcal{F}, \mathcal{G}) \label{same}
\end{eqnarray}
\end{lemma}

{\it Proof:} From (\ref{privacy}), we know that $Q_{\mathcal{T}}^{[k]} \sim Q_{\mathcal{T}}^{[k']}$. Combining with (\ref{query_det}), we have
\begin{eqnarray}
H(Q_{\mathcal{T}}^{[\theta]} | \mathcal{F}) = 0 \label{query_theta}
\end{eqnarray}
From (\ref{indep}), we have
\begin{eqnarray}
&&I(\theta; W_1, \cdots, W_K, \mathcal{F}, \mathcal{G}) = 0 \\
&\overset{(\ref{query_theta})}{\Longrightarrow}& I(\theta; W_1, \cdots, W_K, \mathcal{F}, \mathcal{G}, Q_{\mathcal{T}}^{[\theta]}) = 0 \\
&\overset{(\ref{storage_size})(\ref{query_det})(\ref{answer_det})}{\Longrightarrow}& I(\theta; W_1, \cdots, W_K, \mathcal{F}, \mathcal{G}, A_{\mathcal{T}}^{[\theta]}) = 0 \\
&\overset{}{\Longrightarrow}& (A_{\mathcal{T}}^{[k]}, W_1, \cdots, W_K, \mathcal{F}, \mathcal{G}) \sim (A_{\mathcal{T}}^{[k']}, W_1, \cdots, W_K, \mathcal{F}, \mathcal{G})
\end{eqnarray}
\hfill\QED

\begin{lemma}
For all $\mathcal{K}_c = \{n_1, n_2, \cdots, n_{K_c}\} \subset [1:N]$,
\begin{eqnarray}
H(A_{\mathcal{K}_c}^{[1]} | W_1, \mathcal{F}, \mathcal{G}) = \sum_{n \in \mathcal{K}_c} H(A_n^{[1]} | W_1, \mathcal{F}, \mathcal{G}) \label{indep_answer}
\end{eqnarray}
\end{lemma}
{\it Proof:} From (\ref{storage_size}) and (\ref{mds_property}), we know that for any $K_c$ servers, the stored information is independent. 
\begin{eqnarray}
&&H(W_{k\mathcal{K}_c}) = \sum_{n \in \mathcal{K}_c} H(W_{k n}), \forall k \in [1:K] \\
&\overset{(\ref{indep})}{\Longrightarrow}& H(W_{2\mathcal{K}_c}, \cdots, W_{K\mathcal{K}_c} | W_1, \mathcal{F}, \mathcal{G})
= \sum_{n \in \mathcal{K}_c} \sum_{k=2}^K H(W_{k n} | W_1, \mathcal{F}, \mathcal{G}) \label{m0}
\end{eqnarray}
As answers are functions of the storage, the answers from any $K_c$ servers are independent as well. Consider two arbitrary  subsets of $\mathcal{K}_c$ that have no overlap, $\mathcal{K}_1, \mathcal{K}_2 \subset \mathcal{K}_c, \mathcal{K}_1 \cap \mathcal{K}_2 = \emptyset$.
\begin{eqnarray}
&&I(A_{\mathcal{K}_1}^{[1]}; A_{\mathcal{K}_2}^{[1]} | W_1, \mathcal{F}, \mathcal{G}) \notag\\
&\leq& I(A_{\mathcal{K}_1}^{[1]}; A_{\mathcal{K}_2}^{[1]}, W_{2\mathcal{K}_2}, \cdots, W_{K\mathcal{K}_2} | W_1, \mathcal{F}, \mathcal{G}) \\
&\overset{(\ref{query_det})(\ref{answer_det})}{=}& I(A_{\mathcal{K}_1}^{[1]}; W_{2\mathcal{K}_2}, \cdots, W_{K\mathcal{K}_2} | W_1, \mathcal{F}, \mathcal{G}) \label{eq:awind}\\
&\overset{(\ref{query_det})(\ref{answer_det})}{\leq}& I(W_{2\mathcal{K}_1}, \cdots, W_{K\mathcal{K}_1} ; W_{2\mathcal{K}_2}, \cdots, W_{K\mathcal{K}_2} | W_1, \mathcal{F}, \mathcal{G}) \\
&\overset{(\ref{m0})}{=}& 0 \label{a0}
\end{eqnarray}
Using (\ref{a0}) repeatedly, we obtain (\ref{indep_answer}).
\hfill\QED

Next we proceed to the two scenarios.
To highlight the parameter $K$, in this section, the capacity $C$ and the download cost $D$ are denoted as $C(K)$ and $D(K)$, respectively.
\subsubsection{$(K, N, T, K_c) = (K, 4, 2, 2)$}\label{sec:converse1}
For the setting with $(K, N, T, K_c) = (K, 4, 2, 2)$, we obtain a recursive upper bound that holds for arbitrary $K$. This result is stated in the following theorem.
\begin{theorem}\label{thm:asym1}
For the class of MDS-TPIR instances with $(K,N, T, K_c)=(K,4,2,2)$, with arbitrary $K$, the following recursive relation on the capacity outer bound $\overline{C}(K) \geq C(K)$ holds. 
\begin{eqnarray}
\overline{C}(K) &\overset{}{\leq}& \left( 1 + \frac{3}{8} \left( \frac{1}{\overline{C}(K-1)} \right)  + \left(1 - \left(\frac{2}{3}\right)^{K-1}\right) \frac{3}{4} \right)^{-1}, \forall K \geq 2 \notag\\
\overline{C}{(1)} &=& 1 \label{eq:rec_cap}
\end{eqnarray}
\end{theorem}
\proof Consider an MDS-TPIR instance with $(K,N, T, K_c)=(K,4,2,2)$. When $K = 1$, $\overline{C}{(1)} = 1$ is a trivial bound on $C(1)$. Next we consider $K \geq 2$. Define
\begin{eqnarray}
\overline{C}(K) &=& L/H(A_{1:4}^{[1]} | \mathcal{F}, \mathcal{G}) \label{eq:ck}\\
\overline{C}(K-1) &=& L/H(A_{1:4}^{[2]} | W_1, \mathcal{F}, \mathcal{G}) \label{eq:ck1}
\end{eqnarray}
$\overline{C}(K)$ is a valid outer bound on $C(K)$, since
\begin{eqnarray}
\overline{C}(K) = L/H(A_{1:4}^{[1]} | \mathcal{F}, \mathcal{G}) \geq L/D(K) = C(K)
\end{eqnarray}
Similarly, $\overline{C}(K-1)$ is a valid outer bound on $C(K-1)$.
Now, substituting (\ref{eq:ck}) and (\ref{eq:ck1}) to (\ref{eq:rec_cap}), we have
\begin{eqnarray}
H(A_{1:4}^{[1]} | \mathcal{F}, \mathcal{G})/L \overset{}{\geq} 1 + \frac{3}{8}  H(A_{1:4}^{[2]} | W_{1}, \mathcal{F}, \mathcal{G})/L + \left(1 - \left(\frac{2}{3}\right)^{K-1}\right) \frac{3}{4} \label{eq:rec}
\end{eqnarray}

We proceed to prove (\ref{eq:rec}). To simplify the notation, we define $(W_{1i}, W_{2i}, \cdots, W_{Ki}) = W_{*i}, i \in [1:N]$.
{\begin{eqnarray}
&&H(A_{1:4}^{[1]} | \mathcal{F}, \mathcal{G}) \notag\\
&\overset{(\ref{corr})}{=}& H(A_{1:4}^{[1]}, W_1 | \mathcal{F}, \mathcal{G}) + o(L) L \\
&\overset{(\ref{indep})}{=}& H(W_1) + H(A_{1}^{[1]} | W_{1}, \mathcal{F}, \mathcal{G}) + H(A_{2:4}^{[1]} | W_1, A_{1}^{[1]}, \mathcal{F}, \mathcal{G})  + o(L) L\\
&\overset{}{\geq}& H(W_1) + H(A_{1}^{[1]} | W_{1}, \mathcal{F}, \mathcal{G}) + H(A_{3:4}^{[1]} | W_1, W_{*1}, A_{1}^{[1]}, \mathcal{F}, \mathcal{G}) + o(L) L\\
&\overset{(\ref{h2})(\ref{storage_size})(\ref{answer_det})}{=}& L + H(A_{1}^{[1]} | W_{1}, \mathcal{F}, \mathcal{G}) + H(A_{3:4}^{[1]} | W_1, W_{*1}, \mathcal{F}, \mathcal{G}) + o(L) L\\
&\overset{(\ref{storage_size})(\ref{same})}{=}& L + H(A_{1}^{[2]} | W_{1}, \mathcal{F}, \mathcal{G}) + H(A_{3:4}^{[2]} | W_1, W_{*1}, \mathcal{F}, \mathcal{G})  + o(L) L \label{eq:db_index}
\end{eqnarray}
Advancing the databases indices, from (\ref{eq:db_index}), we have
\begin{eqnarray}
&&H(A_{1:4}^{[1]} | \mathcal{F}, \mathcal{G})  \notag\\
&\geq& L + H(A_{1}^{[2]} | W_{1}, \mathcal{F}, \mathcal{G}) + H(A_{2:3}^{[2]} | W_1, W_{*1}, \mathcal{F}, \mathcal{G}) + o(L) L \label{eq:index2}
\end{eqnarray}
Adding (\ref{eq:db_index}) and (\ref{eq:index2}), we have
\begin{eqnarray}
&&H(A_{1:4}^{[1]} | \mathcal{F}, \mathcal{G}) + o(L) L \notag\\
&\geq& L + H(A_{1}^{[2]} | W_{1}, \mathcal{F}, \mathcal{G}) + \frac{1}{2} \left( H(A_{3:4}^{[2]} | W_1, W_{*1}, \mathcal{F}, \mathcal{G}) + H(A_{2:3}^{[2]} | W_1, W_{*1}, \mathcal{F}, \mathcal{G}) \right) \\
&\overset{}{\geq}& L + H(A_{1}^{[2]} | W_{1}, \mathcal{F}, \mathcal{G}) + \frac{1}{2} \left( H(A_{2:4}^{[2]} | W_1, W_{*1}, \mathcal{F}, \mathcal{G}) + H(A_{3}^{[2]} |  W_1, W_{*1}, \mathcal{F}, \mathcal{G}) \right) \label{eq:sub} \\
&\overset{(\ref{storage_size})(\ref{mds_property})(\ref{eq:awind})}{=}& L + H(A_{1}^{[2]} | W_{1}, \mathcal{F}, \mathcal{G}) + \frac{1}{2} H(A_{3}^{[2]} | W_1, \mathcal{F}, \mathcal{G}) + \frac{1}{2} H(A_{2:4}^{[2]} | W_1, W_{*1}, \mathcal{F}, \mathcal{G}) \label{eq:e1}
\end{eqnarray}
where we use the sub-modular property of entropy functions to obtain (\ref{eq:sub}).
Now consider the term $H(A_{2:4}^{[2]} | W_1, W_{*1}, \mathcal{F}, \mathcal{G})$. This corresponds to the total download for the setting where we have $3$ servers (servers 2, 3 and 4), $K-1$ messages ($W_2, W_3, \cdots, W_K$), each message is of length $L/2$ and the MDS code is fully replicated (conditioning on $W_{*1}$, each other server contains the other half information of entropy $L/2$ about each message), i.e., the TPIR setting. $W_{2}$ is the desired message. As the capacity of this TPIR setting is $\frac{1}{3}\left(1 - \left(\frac{2}{3}\right)^{K-1}\right)^{-1}$ \cite{Sun_Jafar_TPIR}, we have
\begin{eqnarray}
H(A_{2:4}^{[2]} | W_1, W_{*1}, \mathcal{F}, \mathcal{G}) \geq 3\left(1 - \left(\frac{2}{3}\right)^{K-1}\right) \frac{L}{2} \label{eq:tpir}
\end{eqnarray}
Substituting back to (\ref{eq:e1}) and advancing database indices, we have $\forall i, j \in [1:4], i \neq j$,
\begin{eqnarray}
&&H(A_{1:4}^{[1]} | \mathcal{F}, \mathcal{G}) + o(L) L \notag\\
&\overset{}{\geq}& L + H(A_{i}^{[2]} | W_{1}, \mathcal{F}, \mathcal{G}) + \frac{1}{2} H(A_{j}^{[2]} | W_{1}, \mathcal{F}, \mathcal{G}) + \left(1 - \left(\frac{2}{3}\right)^{K-1}\right) \frac{3L}{4} \label{eq:e2}
\end{eqnarray}
Adding (\ref{eq:e2}) for all $i, j \in [1:4]$, we have
\begin{eqnarray}
&&H(A_{1:4}^{[1]} | \mathcal{F}, \mathcal{G}) + o(L) L \notag \\
&\overset{}{\geq}& L + \frac{1}{4} \sum_{i=1}^4 H(A_{i}^{[2]} | W_{1}, \mathcal{F}, \mathcal{G}) + \frac{1}{8} \sum_{j=1}^4 H(A_{j}^{[2]} | W_{1}, \mathcal{F}, \mathcal{G}) + \left(1 - \left(\frac{2}{3}\right)^{K-1}\right) \frac{3L}{4} \\
&\overset{}{\geq}& L + \frac{3}{8}  H(A_{1:4}^{[2]} | W_{1}, \mathcal{F}, \mathcal{G}) + \left(1 - \left(\frac{2}{3}\right)^{K-1}\right) \frac{3L}{4}
\end{eqnarray}
Normalizing both sides by $L$, we arrive at (\ref{eq:rec}).\hfill\QED

Two observations from the converse argument are listed below.
\begin{enumerate}
\item When we set $K=2$, we obtain the information theoretic bound $8/13$.
\begin{eqnarray}
C(2) &\leq& \overline{C}(2) \\
&\overset{(\ref{eq:rec_cap})}{\leq}& (1 + 3/8 \times 1/\overline{C}(1) +  (1-2/3) \times 3/4)^{-1}\\
&\overset{(\ref{eq:rec_cap})}{=}& (1 + 3/8 \times 1 +  (1-2/3) \times 3/4)^{-1} = 8/13 \label{eq:8/13}
\end{eqnarray}

\item As $K\rightarrow\infty$, the capacity upper bound converges to $5/14$.
Since the MDS-TPIR scheme of Freij-Hollanti et al. \cite{FREIJ_HOLLANTI} achieves the rate $1/4$ for this setting as $K\rightarrow \infty$, we note that the asymptotic optimality of the scheme remains open.
\end{enumerate}

\subsubsection{$(K, N, T, K_c)$ with $N < T + K_c$}\label{sec:converse2}
For the setting with $(K, N, T, K_c)$ and $N < T + K_c$, we obtain a recursive upper bound that holds for arbitrary $K$. This result is stated in the following theorem.
\begin{theorem}\label{thm:asym2}
For the class of MDS-TPIR instances $(K,N, T, K_c)$ such that $N < T + K_c$, with arbitrary $K, N, T, K_c$, the following recursive relation on the capacity outer bound $\overline{C}(K) \geq C(K)$ holds. 
\begin{eqnarray}
\overline{C}(K) &\leq& \left( 1 + \frac{N-T}{N} \left( \frac{1}{\overline{C}(K-1)} \right) + (K-1)\left(1-\frac{N-T}{K_c}\right) \right)^{-1}, \forall K \geq 2\notag\\
\overline{C}{(1)} &=& 1
\label{eq:rec_cap_n}
\end{eqnarray}
Therefore, for constant $N, T, K_c$, when $K \rightarrow \infty$, $C(K)$ decreases linearly with $K$ such that downloading everything (rate $1/K$) is order optimal.
\end{theorem}
\proof
Consider an MDS-TPIR instance $(K,N, T, K_c)$ such that $N < T + K_c$. When $K = 1$, $\overline{C}{(1)} = 1$ is a trivial bound on $C(1)$. Next we consider $K \geq 2$. Define
\begin{eqnarray}
\overline{C}(K) &=& L/H(A_{1:N}^{[1]} | \mathcal{F}, \mathcal{G}) \label{eq:ck_n}\\
\overline{C}(K-1) &=& L/H(A_{1:N}^{[2]} | W_1, \mathcal{F}, \mathcal{G}) \label{eq:ck1_n}
\end{eqnarray}
$\overline{C}(K)$ is a valid outer bound on $C(K)$, since
\begin{eqnarray}
\overline{C}(K) = L/H(A_{1:N}^{[1]} | \mathcal{F}, \mathcal{G}) \geq L/D(K) = C(K)
\end{eqnarray}
Similarly, $\overline{C}(K-1)$ is a valid outer bound on $C(K-1)$.
Now, substituting (\ref{eq:ck_n}) and (\ref{eq:ck1_n}) to (\ref{eq:rec_cap_n}), we have
\begin{eqnarray}
\frac{H(A_{1:N}^{[1]} | \mathcal{F}, \mathcal{G})}{L} \geq 1 + \frac{N-T}{N} \frac{H(A_{1:N}^{[2]} | W_1, \mathcal{F}, \mathcal{G})}{L} + (K-1)\left(1-\frac{N-T}{K_c}\right)
\label{eq:rec_n}
\end{eqnarray}

We proceed to prove (\ref{eq:rec_n}). 
Consider an index set $\mathcal{N} \subset [1:N]$ with cardinality $|\mathcal{N}| = N-T  < K_c$. Denote the complement of $\mathcal{N}$ as $\mathcal{N}^c$.
\begin{eqnarray}
&&%D(K) \geq 
H(A_{1:N}^{[1]} | \mathcal{F}, \mathcal{G}) \notag\\
&\overset{(\ref{corr})}{=}& H(A_{1:N}^{[1]}, W_1 | \mathcal{F}, \mathcal{G}) +o(L) L \\
&\overset{(\ref{indep})}{=}& H(W_1) + H(A_{\mathcal{N}}^{[1]} | W_1, \mathcal{F}, \mathcal{G}) + H(A_{\mathcal{N}^c}^{[1]} | W_1, A_{\mathcal{N}}^{[1]}, \mathcal{F}, \mathcal{G}) +o(L) L \\
&\overset{(\ref{h2})(\ref{indep_answer})}{\geq}& L + \sum_{n \in \mathcal{N}} H(A_n^{[1]} | W_1, \mathcal{F}, \mathcal{G}) 
+ H(A_{\mathcal{N}^c}^{[1]} | W_1, W_{*\mathcal{N}}, A_{\mathcal{N}}^{[1]}, \mathcal{F}, \mathcal{G}) +o(L) L\\
&\overset{(\ref{storage_size})(\ref{query_det})(\ref{answer_det})}{=}& L + \sum_{n \in \mathcal{N}} H(A_n^{[1]} | W_1, \mathcal{F}, \mathcal{G})  + H(A_{\mathcal{N}^c}^{[1]} | W_1, W_{*\mathcal{N}}, \mathcal{F}, \mathcal{G}) +o(L) L \label{eq:s1}\\
&\overset{(\ref{storage_size})(\ref{same})}{=}& L + \sum_{n \in \mathcal{N}} H(A_n^{[2]} | W_1, \mathcal{F}, \mathcal{G})  + H(A_{\mathcal{N}^c}^{[2]} | W_1, W_{*\mathcal{N}}, \mathcal{F}, \mathcal{G}) +o(L) L \\
&\overset{(\ref{storage_size})(\ref{query_det})(\ref{answer_det})}{=}& L + \sum_{n \in \mathcal{N}} H(A_n^{[2]} | W_1, \mathcal{F}, \mathcal{G}) + H(A_{{1:N}}^{[2]} | W_1, W_{*\mathcal{N}}, \mathcal{F}, \mathcal{G}) +o(L) L \\
&\overset{(\ref{corr})}{\geq}& L + \sum_{n \in \mathcal{N}} H(A_n^{[2]} | W_1, \mathcal{F}, \mathcal{G}) + H(A_{1:N}^{[2]}, W_2 | W_1, W_{*\mathcal{N}}, \mathcal{F}, \mathcal{G})  + o(L) L \\
&\overset{}{\geq}& L + \sum_{n \in \mathcal{N}} H(A_n^{[2]} | W_1, \mathcal{F}, \mathcal{G}) + H(W_2 | W_1, W_{*\mathcal{N}}, \mathcal{F}, \mathcal{G}) +  H(A_{1:N}^{[2]} | W_1, W_2, W_{*\mathcal{N}}, \mathcal{F}, \mathcal{G}) + o(L) L\notag\\
&&\\
&\overset{(\ref{indep})(\ref{storage_size})(\ref{mds_property})}{=}& L + \sum_{n \in \mathcal{N}} H(A_n^{[2]} | W_1, \mathcal{F}, \mathcal{G}) + L(K_c - |\mathcal{N}|)/K_c + H(A_{1:N}^{[2]} | W_1, W_2, W_{*\mathcal{N}}, \mathcal{F}, \mathcal{G})+ o(L) L \label{eq:uselater}\\
&\overset{(\ref{storage_size})(\ref{query_det})(\ref{answer_det})}{=}& L + \sum_{n \in \mathcal{N}} H(A_n^{[2]} | W_1, \mathcal{F}, \mathcal{G}) + L(K_c - N + T)/K_c + H(A_{\mathcal{N}^c}^{[2]} | W_1, W_2, W_{*\mathcal{N}}, \mathcal{F}, \mathcal{G}) + o(L) L\notag\\
&&\label{eq:s2}
\end{eqnarray}
To bound the term $H(A_{\mathcal{N}^c}^{[2]} | W_1, W_2, W_{*\mathcal{N}}, \mathcal{F}, \mathcal{G})$, we repeat (\ref{eq:s1}) to (\ref{eq:s2}) for messages $W_3, \cdots, W_K$. This gives us
\begin{eqnarray}
&&H(A_{1:N}^{[1]} | \mathcal{F}, \mathcal{G}) \notag\\
&\geq& L + \sum_{n \in \mathcal{N}} H(A_n^{[2]} | W_1, \mathcal{F}, \mathcal{G}) + L(K-1)\left(1-\frac{N-T}{K_c}\right) + o(L) L \label{eq:s3}
\end{eqnarray}
Consider (\ref{eq:s3}) for all subsets of $[1:N]$ that have exactly $N-T$ elements and average over all such subsets. We have
\begin{eqnarray}
&&H(A_{1:N}^{[1]} | \mathcal{F}, \mathcal{G}) \notag\\
&\geq& L + \frac{1}{\binom{N}{N-T}} \sum_{\mathcal{N}: |\mathcal{N}| = N -T} \sum_{n \in \mathcal{N}} H(A_n^{[2]} | W_1, \mathcal{F}, \mathcal{G}) + L(K-1)\left(1-\frac{N-T}{K_c}\right) + o(L) L \\
&\geq& L + \frac{N-T}{N} H(A_{1:N}^{[2]} | W_1, \mathcal{F}, \mathcal{G}) + L(K-1)\left(1-\frac{N-T}{K_c}\right) + o(L) L
\end{eqnarray}
Letting $L \rightarrow \infty$} and normalizing by $L$, we have proved (\ref{eq:rec_n}) and (\ref{eq:rec_cap_n}). \hfill\QED

Based on Theorem \ref{thm:asym2} the following observations are relevant.
\begin{enumerate}
\item When we set $K=2, K_c = N-1$, we obtain the information theoretic bound for Theorem \ref{thm:class}, i.e., $(N^2 - N)/(2N^2 - 3N + T)$.
\begin{eqnarray}
C(2) &\leq& \overline{C}(2) \\
&\overset{(\ref{eq:rec_cap_n})}{\leq}& \left(1 + \frac{N-T}{N} \times \frac{1}{\overline{C}(1)} +  (2-1)\left(1-\frac{N-T}{N-1}\right) \right)^{-1}\\
&\overset{(\ref{eq:rec_cap_n})}{=}& \left(1 + \frac{N-T}{N} \times 1 +  \frac{T-1}{N-1}\right)^{-1} = \frac{N^2 - N}{2N^2 - 3N + T}
\end{eqnarray}
\item As $K\rightarrow\infty$, Theorem \ref{thm:asym2} shows that the capacity decays as $1/K$, so that it converges to $0$. As a sanity check, we note that indeed, the MDS-TPIR scheme of Freij-Hollanti et al. \cite{FREIJ_HOLLANTI}, which does not depend on the number of messages $K$, does not apply when $N<T+K_c$. Thus, in this case the asymptotic optimality as $K\rightarrow \infty$ is trivially settled.
\end{enumerate}

\subsection{Restricted Colluding Sets}\label{sec:restricted}
Recall that for the setting of our counterexample, i.e., $(K, N, T, K_c)=(2,4,2,2)$, while the linear capacity is settled, the information theoretic capacity remains open. In particular, the best information theoretic capacity upper bound that we were able to obtain is $8/13$. To gain  insights into the potential tightness of this bound, here we look into the capacity of this setting with restricted colluding sets, a line of inquiry recently initiated by Tajeddine et al. in \cite{Tajeddine_Gnilke_Karpuk}. Our motivation for studying restricted colluding sets comes from the following observation.

Consider TPIR, for which the capacity is known \cite{Sun_Jafar_TPIR}. The TPIR formulation allows the possibility that \emph{any} set of up to $T$ servers may collude. However, suppose we relax the privacy constraint, by allowing only collusions between cyclically contiguous servers, i.e., the colluding servers must belong to the set of servers indexed $\{n, n+1, \cdots, n+T-1\}$ for some $n\in[1:N]$, with the indices interpreted modulo $N$. Because of the symmetry that is still maintained across servers, it is readily verified that the converse proof for TPIR in \cite{Sun_Jafar_TPIR} still goes through unchanged. Thus, even though the restriction on colluding sets to cyclically contiguous servers relaxes the privacy constraint, it does not affect the capacity of TPIR.

This  leads us to question if a similar property might hold for MDS-TPIR. If so, then we could gain insights into the capacity of MDS-TPIR by imposing similar restrictions on the colluding sets. This line of thought leads us to two somewhat contrasting observations, that are presented in the following two subsections.

\subsubsection{$(K, N, T,K_c) = (2,4,2,2)$ with Cyclically Adjacent Colluding Sets}\label{sec:restricted1}
Our first observation is in favor of the tightness of the upper bound $8/13$. Indeed, if colluding sets were restricted to cyclically contiguous sets then $8/13$ is the capacity for the MDS-TPIR setting $(K, N, T, K_c)=(2,4,2,2)$. This observation is summarized in a bit more detail next.

For our counterexample we  considered the MDS-TPIR setting $(K, N, T, K_c) = (2,4,2,2)$ where any 2 servers may collude. Suppose, now we restrict the colluding sets of servers to cyclically adjacent pairs, i.e., any one of $\{1,2\}, \{2,3\}, \{3,4\}, \{4,1\}$.  Essentially we have relaxed the privacy constraint by eliminating the possibilities that Server $1$ might collude with Server $3$, or that Server $2$ might collude with Server $4$. For this setting, we show that the capacity is $8/13$.

%(put in a theorem or not?)

The converse is similar to that with $T = 2$, presented in Section \ref{sec:converse1}. (\ref{eq:tpir}) holds with restricted colluding sets when $K = 2$, because we are left with only $K - 1 = 1$ message. All other steps follow similarly because the assumption of symmetry across servers  holds under cyclically adjacent colluding sets. As a result, the capacity upper bound of $8/13$ (refer to (\ref{eq:8/13})) holds here.

Next, we summarize the achievable scheme. The message construction and the storage code are specified as follows. 
\begin{eqnarray}
&&W_{kn} \in \mathbb{F}_p^{4\times 1}, k \in [1:2], n \in [1:4] \\
&&W_{k} = (W_{k1}; W_{k2}) \in \mathbb{F}_p^{8 \times 1} \\
&&W_{k3} = W_{k1} + W_{k2}, W_{k4} = W_{k1} + 2W_{k2}
\end{eqnarray}
The construction of queries is similar to that with $T= 2$ in Section \ref{sec:main}. The query to each server $Q_n^{[k]}$ is comprised of two parts, $Q_n^{[k]}(W_1), Q_n^{[k]}(W_2)$. Each part contains $2$ row vectors, along which the server should project its corresponding stored message symbols. To generate the query vectors, the user privately chooses two matrices, $S = (V_1;V_2; V_3; V_4)$ and $S' = (U_0; U_1; U_2; U_3)$, independently and uniformly from $\mathcal{S}_4$, the set of all full rank $4 \times 4$ matrices over $\mathbb{F}_p$. Define
\begin{eqnarray}
\mathcal{V}_1=\{V_1, V_2\},&&\mathcal{U}_1=\{U_0, U_1+U_2\}\\
\mathcal{V}_2=\{V_2, V_3\},&&\mathcal{U}_2=\{U_0, U_1+2U_2\}\\
\mathcal{V}_3=\{V_3, V_4\},&&\mathcal{U}_3=\{U_0, U_1\}\\
\mathcal{V}_4=\{V_4, V_1\},&&\mathcal{U}_4=\{U_0, U_2\}
\end{eqnarray}
Independent random orderings of the rows in $\mathcal{V}_n$ are the queries to Server $n$ for the desired message and independent random orderings of the rows in $\mathcal{U}_n$ are the queries to Server $n$ for the undesired message. The rate achieved is $8/13$ because the $8$ desired symbols along the $V_i$ vectors are all independent and the $8$ undesired symbols occupy only $5$ dimensions (the $4$ symbols along $U_0$ contribute only $2$ independent dimensions and the remaining $4$ symbols contribute only $3$ independent dimensions). Privacy follows from the observation that for each cyclically adjacent colluding set of servers, say Server $1$ and Server $2$, the sets $\mathcal{V}_1,\mathcal{V}_2$ intersect in one of their elements, as do the sets $\mathcal{U}_1,\mathcal{U}_2$, and both are otherwise uniformly random, thus making the distinction of $\mathcal{U},\mathcal{V}$ invisible to the colluding servers. Note that this scheme is not private to the non-adjacent colluding servers, say Server $1$ and Server $3$, because, $\mathcal{V}_1,\mathcal{V}_3$ contain no common vectors, while $\mathcal{U}_1,\mathcal{U}_3$ do share a common vector. The remaining details are virtually identical to the settings already covered in Section \ref{sec:main} and Section \ref{sec:ach_proof2} and are omitted.

%\bigskip
%Now let us compare this result with restricted colluding sets with that of the $T = 2$ setting in Section \ref{sec:main}, where we know that $8/13$ is an outer bound, but we could not achieve the rate $8/13$ if any $T= 2$ servers may collude so that we do not know if $8/13$ is the capacity.

%Note that we have showed that 8/13 can be achieved if the colluding sets are disjoint, i.e., if only the first $T=2$ servers may collude and the last $T=2$ servers may collude, because these disjoint colluding sets are contained in (\ref{eq:collude_set}). For TPIR, when $N$ is a multiple of $T$, we know that the capacity with disjoint colluding sets, where are colluding sets are restricted to be disjoint sets of $T$ servers, is the same as the capacity of TPIR \cite{Sun_Jafar_TPIR}, where any $T$ servers could collude. So, it might be tempting to use that intuition to conjecture that $8/13$ might be the capacity for MDS-TPIR with $(K, N, T, K_c) = (2, 4, 2, 2)$. But that intuition does not carry to MDS-TPIR, as we show next. In particular, we show that the capacity for disjoint colluding sets with cardinality $T$ can be strictly higher than the capacity of all colluding sets with cardinaltiy $T$ for MDS-TPIR in general. Therefore, for the MDS-TPIR setting with $(K, N, T, K_c) = (2, 4, 2, 2)$, even if $8/13$ is the capacity for disjoint colluding sets (in fact, the capacity for disjoint colluding sets is $2/3$, as shown in Section \ref{sec:ex}), $8/13$ may still not be the capacity when all colluding sets are allowed.

%\subsection{Disjoint Colluding Sets of $T$ Servers Each}
\subsubsection{Disjoint Colluding Sets of $T$ Servers Each}
Our second observation provides a counterpoint to the first observation. The first observation favored the tightness of $8/13$ bound based on the insight originating from TPIR, that certain restrictions on colluding sets may not affect capacity. The second observation challenges this viewpoint by showing that insights from TPIR do not carry over to MDS-TPIR.

Consider again the TPIR problem. Suppose $T$ divides $N$, i.e., $mT=N$ for some $m\in\mathbb{Z}_+$, and we partition the $N$ servers into the $m$ disjoint sets of $T$ elements each: $\mathcal{T}_1=\{1,2,\cdots, T\}$, $\mathcal{T}_2=\{T+1,T+2,\cdots, 2T\}$, $\cdots$, $\mathcal{T}_m=\{(m-1)T+1, (m-1)T+2, \cdots, N\}$. Further, suppose we relax the privacy constraint and allow collusions between only those servers that belong to the same $\mathcal{T}_i$, $i\in[1:m]$. Then, note that the TPIR problem with restricted colluding sets becomes equivalent to the PIR problem with $N/T=m$ servers.\footnote{This is because storage is fully replicated, so that each disjoint set of $T$ colluding servers may be equivalently replaced with $1$ server.} However, the capacity of PIR with $N/T$ servers is the \emph{same} as the capacity of TPIR with $N$ servers. Therefore, relaxing the privacy constraint by restricting the colluding sets to disjoint sets of cardinality $T$ each, in the manner described above, does not affect the capacity of TPIR. However, as we will show next, the same is not true for MDS-TPIR.

Consider MDS-TPIR with $(K, N, T, K_c)=(2,4,3,2)$, where any $T=2$ of the $N=4$ servers may collude. From Theorem \ref{thm:class} we know that the capacity of this setting is $6/11$. 
However, now suppose we partition the servers into disjoint sets $\mathcal{T}_1=\{1,2\}$, $\mathcal{T}_2=\{3,4\}$, each of cardinality $T=2$. Now we allow collusions only between servers in the same $\mathcal{T}_i$ set, i.e., Server $1$ can only collude with Server $2$, while Server $3$ can only collude with Server $4$. Then, in contrast to TPIR where such a restriction on colluding sets does not affect the capacity, we now show that with these restricted colluding sets, the capacity of MDS-TPIR changes --- it increases from $6/11$ to $4/7$.

The converse for rate $4/7$ is trivial, because the rate can not be higher than that of MDS-PIR with $(K, N, K_c) = (2,4,3)$, where privacy needs to be ensured only to each individual server. From \cite{Banawan_Ulukus}, we know that the capacity of MDS-PIR with $(K, N, K_c) = (2,4,3)$ is $4/7$. Therefore, the upper bound follows.

Next, we consider the achievable scheme. Each message consists of $12$ symbols. The storage code is specified as follows. 
\begin{eqnarray}
&&W_{kn} \in \mathbb{F}_p^{4\times 1}, k \in [1:2], n \in [1:4] \\
&&W_{k} = (W_{k1}; W_{k2}; W_{k3}) \in \mathbb{F}_p^{12 \times 1} \\
&&W_{k4} = W_{k1} + W_{k2} + W_{k3}
\end{eqnarray}
The query to each server $Q_n^{[k]}$ is comprised of vectors in $\mathcal{V}_n$ and $\mathcal{U}_n$, given as follows. 
\begin{eqnarray}
\mathcal{V}_1=\{V_1, V_3, V_5\},&&\mathcal{U}_1=\{U_0, U_1, U_2\}\\
\mathcal{V}_2=\{V_1, V_3, V_5\},&&\mathcal{U}_2=\{U_0, U_1, U_2\}\\
\mathcal{V}_3=\{V_2, V_4, V_6\},&&\mathcal{U}_3=\{U_0, U_1, U_2\}\\
\mathcal{V}_4=\{V_2, V_4, V_6\},&&\mathcal{U}_4=\{U_0, U_1,U_2\}
\end{eqnarray}
where $S = (V_1;V_2; V_3; V_4; V_5; V_6)$ and $S' = (U_0; U_1; U_2; U_3; U_4; U_5)$ are independent and uniform from the set of all full rank $6 \times 6$ matrices. 
The rate achieved is $12/(12+9) = 4/7$ because the $12$ desired symbols along the $V_i$ vectors are all independent and the $12$ undesired symbols occupy only $9$ dimensions (the symbols along each $U_i$, $i\in\{0,1,2\}$, occupy only $K_c = 3$ dimensions). Privacy follows from the observation that for either colluding set $\{1,2\}$ or $\{3,4\}$, the vectors in $\mathcal{V}$ and $\mathcal{U}$ are both the same. The remaining details can be filled in based on Section \ref{sec:main} and Section \ref{sec:ach_proof2} and are omitted. 

In light of the two contrasting observations, the tightness of the $8/13$ upper bound, as well as the general impact of restricted colluding sets on the capacity of MDS-TPIR remain intriguing open problems for future work. For readers interested in the latter problem, we conclude this section with two simple examples of such capacity characterizations. 

\subsubsection{Examples of Capacity of MDS-TPIR under Restricted Colluding Sets}\label{sec:ex}
As usual in this section, we will omit details of achievability arguments that follow directly from Section \ref{sec:main} and Section \ref{sec:ach_proof2}.

\paragraph{Example 1}
Consider the setting $(K, N, K_c)=(2,4,2)$ and let the restricted colluding sets be $\{1,2\}, \{3,4\}$. Alternatively, let the restricted colluding sets be $\{1,2\}, \{3\}, \{4\}$. In either case, the capacity is $2/3$, same as that of MDS-PIR with $(K, N, K_c) = (2, 4, 2)$ \cite{Banawan_Ulukus} so that the converse is implied. The scheme that achieves rate $4/6= 2/3$ is as follows.
\begin{eqnarray}
&&W_{kn} \in \mathbb{F}_p^{2\times 1}, k \in [1:2], n \in [1:4] \\
&&W_{k} = (W_{k1}; W_{k2}) \in \mathbb{F}_p^{4 \times 1} \\
&&W_{k3} = W_{k1} + W_{k2}, W_{k4} = W_{k1} + 2W_{k2}\\
&& \mathcal{V}_1=\{V_1\}, \mathcal{U}_1=\{U_0\}\\
&& \mathcal{V}_2=\{V_1\}, \mathcal{U}_2=\{U_0\}\\
&& \mathcal{V}_3=\{V_2\}, \mathcal{U}_3=\{U_0\}\\
&& \mathcal{V}_4=\{V_2\}, \mathcal{U}_4=\{U_0\}
\end{eqnarray}
where $S = (V_1;V_2)$ and $S' = (U_0; U_1)$ are independently and uniformly chosen from the set of all full rank $2 \times 2$ matrices. 

\paragraph{Example 2}
Suppose $(K, N, K_c) = (2, 3, 2)$ and the colluding sets are either $\{1,2\},\{2,3\}$.
Alternatively, suppose the colluding sets are $\{1,2\}, \{3\}$. In both cases, the capacity is $4/7$. The scheme that achieves rate $4/7$ is as follows.
\begin{eqnarray}
&&W_{kn} \in \mathbb{F}_p^{2\times 1}, k \in [1:2], n \in [1:3] \\
&&W_{k} = (W_{k1}; W_{k2}) \in \mathbb{F}_p^{4 \times 1} \\
&&W_{k3} = W_{k1} + W_{k2}\\
&& \mathcal{V}_1=\{V_1\}, \mathcal{U}_1=\{U_0\}\\
&& \mathcal{V}_2=\{V_1, V_2\}, \mathcal{U}_2=\{U_0, U_1\}\\
&& \mathcal{V}_3=\{V_2\}, \mathcal{U}_3=\{U_0\}
\end{eqnarray}
where $S = (V_1;V_2)$ and $S' = (U_0; U_1)$ are independent and uniformly chosen from the set of all full rank $2 \times 2$ matrices over $\mathbb{F}_p$. 

For the converse, consider (\ref{eq:uselater}). Plugging in $K = 2, K_c = 2, \mathcal{N} = \{3\}, N = 3$, we have
\begin{eqnarray}
D \geq H(A_{1:3}^{[1]} | \mathcal{F}, \mathcal{G}) \geq L + H(A_{3}^{[2]} | W_1, \mathcal{F}, \mathcal{G}) + L/2 + o(L) L 
\end{eqnarray}
Note that (\ref{eq:uselater}) still holds when $|\mathcal{N}| = K_c$. Plugging in $K = 2, K_c = 2, \mathcal{N} = \{1,2\}, N = 3$, we have
\begin{eqnarray}
D \geq H(A_{1:3}^{[1]} | \mathcal{F}, \mathcal{G}) \geq L + H(A_{1}^{[2]} | W_1, \mathcal{F}, \mathcal{G}) + H(A_{2}^{[2]} | W_1, \mathcal{F}, \mathcal{G}) + o(L) L 
\end{eqnarray}
Adding the two inequalities above, we have
\begin{eqnarray}
2D &\geq& 5L/2 +  H(A_{1}^{[2]}, A_{2}^{[2]},A_{3}^{[2]} | W_1, \mathcal{F}, \mathcal{G}) + L/2 + o(L) L  \\
&\overset{(\ref{corr})}{\geq}& 5L/2 +  H(W_2 | W_1, \mathcal{F}, \mathcal{G}) + L/2 + o(L) L  \\
&\overset{(\ref{indep})(\ref{h2})}{=}& 7L/2 + o(L) L
\end{eqnarray}
Normalizing by $L$ and taking limits as $L$ approaches infinity, gives us the upper bound on the rate $L/D$ as $4/7$, which completes the converse.

\bibliographystyle{IEEEtran}
\bibliography{Thesis}
\end{document}